\documentclass[acmtog]{acmart}
\acmSubmissionID{461}

\usepackage{booktabs} 
\usepackage{diagbox}
\usepackage{subfig}
\usepackage{multirow}
\usepackage{mathrsfs}
\usepackage{graphicx}
\usepackage{enumitem}
\usepackage{soul}
\usepackage{url}
\usepackage[normalem]{ulem}
\usepackage[skip=3pt]{caption}
\setcitestyle{square} 

\usepackage[ruled]{algorithm2e} 

\SetAlFnt{\small}
\SetAlCapFnt{\small}
\SetAlCapNameFnt{\small}
\SetAlCapHSkip{0pt}

\newcommand{\new}[1]{\textcolor{black}{#1}} 
\newcommand{\fillin}[1]{\textcolor{black}{#1}}

\setcopyright{rightsretained}
\acmDOI{XXXX} \acmVolume{XX} \acmNumber{XX}

\begin{document}

\author{Valentin Deschaintre*}
\affiliation{%
	\institution{Adobe Research}
	\country{UK}
}
\email{deschain@adobe.com}

\author{Julia Guerrero-Viu*}
\affiliation{%
	\institution{Universidad de Zaragoza - I3A}
	\country{Spain}
}
\email{juliagviu@unizar.es}

\author{Diego Gutierrez}
\affiliation{%
	\institution{Universidad de Zaragoza - I3A}
	\country{Spain}
}
\email{diegog@unizar.es}

\author{Tamy Boubekeur}
\affiliation{%
	\institution{Adobe Research}
	\country{France}
}
\email{boubek@adobe.com}

\author{Belen Masia}
\affiliation{%
	\institution{Universidad de Zaragoza - I3A}
	\country{Spain}
}
\email{bmasia@unizar.es}

\newcommand\blfootnote[1]{%
  \begingroup
  \renewcommand\thefootnote{}\footnote{#1}%
  \addtocounter{footnote}{-1}%
  \endgroup
}

\title{The Visual Language of Fabrics}

\begin{abstract}
\blfootnote{* Joint first authors}
We introduce text2fabric, a novel dataset that links free-text descriptions to various fabric materials. 
The dataset comprises 15,000 natural language descriptions associated to 3,000 corresponding images of fabric materials. 
Traditionally, material descriptions come in the form of tags/keywords, which limits their expressivity, induces pre-existing knowledge of the appropriate vocabulary, and ultimately leads to a chopped description system. 
Therefore, we study the use of free-text as a more appropriate way to describe material appearance, taking the use case of fabrics as a common item that non-experts may often deal with. 
Based on the analysis of the dataset, we identify a compact lexicon, set of attributes and key structure that emerge from the descriptions. 
This allows us to accurately understand how people describe fabrics and draw directions for generalization to other types of materials. 
We also show that our dataset enables specializing large vision-language models such as CLIP, creating a meaningful latent space for fabric appearance, and significantly improving applications such as fine-grained material retrieval and automatic captioning.
 \end{abstract}

\ccsdesc[500]{Computing methodologies~Appearance and texture representations}

\keywords{material appearance, perception, descriptions}

 \begin{teaserfigure}
 \includegraphics[width=\textwidth, height=2.5in]{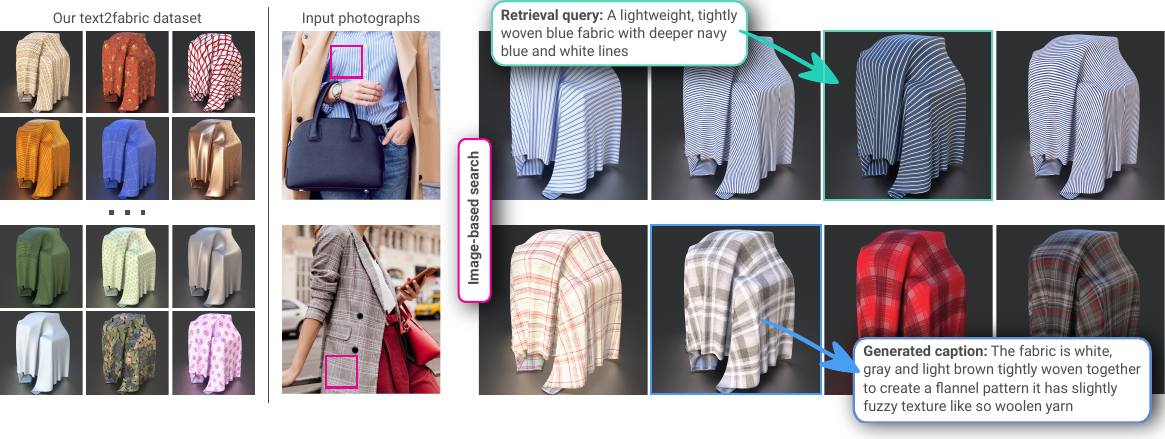}
\caption{
Our text2fabric dataset links high-quality renderings of a large variety of fabric materials 
to natural language descriptions of their appearance. 
We conduct a thorough analysis of this dataset, and leverage it to fine-tune large-scale vision-language models for a variety of tasks.
We show here examples of such tasks: (i) image-based search, even using real photographs as input, yields relevant results from our dataset (the magenta square in the photographs marks the input crop, and the corresponding search results can be found in each row); (ii) text-based queries (green) result in successful fine-grained retrieval within the dataset; and, (iii) given an input image, we can generate detailed and rich descriptions of appearance (blue).
Our work not only derives interesting insights regarding how people describe (fabric) appearance, but also demonstrates that a relatively small amount of high-quality data enables successful application of large vision-language models to specialized domains.}
 \label{fig:teaser}
 \end{teaserfigure}
\maketitle

\newcommand{\absfreq}[0]{f}
\newcommand{\arf}[0]{arf}
\newcommand{\coverage}[0]{cov}
\newcommand{\nattr}[0]{N_a}
\newcommand{\ndescr}[0]{D}
\newcommand{\word}[0]{w}

\newcommand{\nbvaliddescription}{15,461 }
\newcommand{\nbinvaliddescription}{3,706 }
\newcommand{\nbvaliddescribers}{122 }

\newcommand{\nbtokenspostpro}{191,783}
\newcommand{\nbtypespostpro}{3,539 }
\newcommand{\nblemmaspostpro}{2,762 }
\section{Introduction}

The recent quality surge in multimodal natural language processing (NLP) and vision-language models has enabled new interaction possibilities between images and text~\cite{CLIP,BLIP,ramesh2022hierarchical, saharia2022photorealistic,poole2022dreamfusion,chen2022tango}. 
However, the underlying text-to-image models are trained on hundreds of millions of data points collected online, with a bias towards natural images and pictures typically found on the internet, and with general descriptions that do not capture the fine details and rich subtleties that domain-specific applications may require. 

In this paper we explore the use of natural language to convey \textit{fine-grained} material appearance. We pose three open key questions: (i) Is there an underlying common lexicon and structure when people describe material appearance using natural language? (ii) Can we communicate material appearance precisely enough with natural language only? (iii) Are language concepts relevant to material appearance well understood by large foundational models~\cite{CLIP, BLIP}?

Although these questions are relevant for any type of vision-language model dealing with material appearance, to make the task tractable we focus on the particular class of \emph{fabrics}. We choose this class since fabrics exhibit a wide variety of looks, patterns and reflectivity properties at different scales, are familiar to everyone, and are  ubiquitous, widely present in many daily scenarios.

We first build a large dataset, text2fabric, relating photorealistic renderings of digital fabrics covering a wide range of appearance to natural language descriptions provided through crowdsourcing. We then perform a thorough analysis of our data, aimed at improving our understanding of how fabric appearance is described, and find that: (i) there is a common lexicon (ca. 500 words are enough to cover 95\% of the 15,461 valid descriptions gathered); (ii) common properties (attributes) of appearance emerge from the descriptions; (iii) users do follow a certain structure when describing fabric appearance; and (iv) despite the infinite description space provided by natural language, there is a high similarity between descriptions of the same fabric. 
All these findings suggest that we have a shared understanding of language as it relates to material (fabric) appearance, which is key to \new{communicating} appearance precisely.

In addition, we leverage two successful and widely used vision-language models ---CLIP~\cite{CLIP} and BLIP~\cite{BLIP}--- 
and show how their performance improves significantly when fine-tuned on our dataset. 
Last, we demonstrate applications of our model for various tasks (see Figure \ref{fig:teaser}), including fine-grained text-based retrieval, image-based search, and automatic description or captioning of fabrics, 
all of them robust in the presence of light and geometry variations.

In summary, we present the following contributions:
\begin{itemize}
    \item A text2fabric dataset including 15,000+ descriptions associated with 3,000 different fabric materials. The dataset is further augmented with 42,000 additional images featuring different geometries and lighting.
    \item A \new{general methodology to collect and analyze natural language data describing images of fabrics, which is} applicable to other domains. 
    \item The identification of a common lexicon, structure and curated set of attributes that are relevant when describing fabrics.
    \item Fine-tuned models demonstrating the benefit of our dataset on several tasks.
\end{itemize}

Our full text2fabric dataset, as well as the fine-tuned models, are made publicly available to facilitate future research\footnote{Project website: \url{https://valentin.deschaintre.fr/text2fabric}}. 

\section{Related work}

\paragraph{Description of visual attributes} Describing the appearance of objects, scenes or situations through language is a common task for humans. It allows to transmit richer information than simpler labeling and categorization approaches. Descriptions are not only more natural, but they also allow to focus on key or unusual aspects, or to add comparisons or semantics~\cite{farhadi2009describing}. This information would then be leveraged by means of natural language processing (NLP) to enable new, more user-friendly computer graphics and vision algorithms. While providing a complete, general method for \textit{any} object is still a daunting task, several methods have been proposed for people~\cite{bourdev2011describing}, faces~\cite{kumar2011describable}, or scenes~\cite{patterson2012sun}. Other authors have focused on the particular problem of texture description. Bhushan et al.~\shortcite{bhushan1997texture} came up with a limited 98-word lexicon which could describe 82\% of their experimental data, formed by textures; instead of using text descriptions, participants had to cluster words based on perceived similarity. Inspired by this work, Cipoi and colleagues~\shortcite{cimpoi2014describing} introduced the Describable Textures Dataset (DTD), composed of more 5,000 images labeled with one or more adjectives in a simple lexicon of 47 texture terms. The work was later extended into the Describable Textures in Detail Dataset (DTD$^2$)~\cite{wu2020describing}, including natural language descriptions. Recently, Xu et al.~\shortcite{xu2022texture} presented Texture BERT, a learning-based architecture that minimizes distances between texture and text features, optimized for image retrieval. While the domain of texture descriptions is rich and varied, our notion of appearance goes beyond 2D RGB maps, including aspects like reflectance, glossiness, touch, use, weight, etc. 
Our methodology further has the potential to generalize to other material classes.

\paragraph{Perceptually-meaningful material spaces}
The role of perception in computer graphics has been extensively researched \cite{bartz2008role,mcnamara2011perception,fleming2015perception,thompson2011visual}. In particular, finding perceptually-meaningful material spaces has many applications in graphics, including editing \cite{serrano2016intuitive,shi2021}, gamut mapping \cite{sun2017attribute}, image-space manipulations \cite{delanoy2022generative,khan2006image,boyadzhiev2015band} or material similarity \cite{lagunas2019similarity}.

Pellacini et al.~\shortcite{pellacini2000toward} derived a two-dimensional perceptually uniform space for gloss, correlated with the parameteres of the Ward BRDF model \cite{ward1992measuring}; the concept was later extended by Wills and colleagues~\shortcite{wills2009toward} to include different reflectance models. Focusing on the problem of optimal reflectance acquisition, Nielsen et al. presented a perceptual scaling and decomposition of BRDF data, which allowed to reduce PCA dimensionality; the authors further showed how the first few dimensions roughly correlate with the specular and diffuse components of appearance~\cite{nielsen2015optimal}. In computer graphics, the joint effect of geometry and illumination on appearance has also been thoroughly studied ~\cite{lagunas2021joint,bousseau2011optimizing,vangorp2007influence,dror2001estimating,storrs2021unsupervised}. Recently, Serrano and colleagues~\shortcite{serrano2021effect} trained a deep learning architecture using over 40,000 combinations or shape, material and illumination, to predict perceptual attributes of materials that correlate with human judgements.

\paragraph{Distilling human-centered knowledge}
Understanding how people perform certain tasks and interact with different concepts is an important aspect of many human-centered computer graphics applications. Gathering rich, annotated datasets allows to distill this knowledge and apply it to the design of intuitive interfaces and workflows, help the development of novel algorithms, and automate time-consuming tasks. 
For instance, Cole et al.~\shortcite{Cole2008} and Gryaditskaya et al.~\shortcite{gryaditskaya2019opensketch} gathered a dataset of sketches and carefully analysed the practices of artists in terms of line types and how they are used. Garces et al. \shortcite{ garces2014similarity} provided a measure of style similarity for clip art by gathering and analyzing human responses in a dataset of a thousand elements. The \textit{OpenSurfaces} dataset \cite{li2018learning} contains crowdsourced pairwise comparisons of material properties, to improve the performance on difficult tasks such as intrinsic image decomposition. Jarabo and colleagues \shortcite{jarabo2014people} tackled the problem of navigating the four-dimensional structure of light fields to provide an intuitive interface for editing them. Last, data from over 800 participants exploring VR scenes has been used to devise novel compression or video synopsis algorithms \cite{sitzmann2018saliency}.

We frame our data gathering and analysis in a similar fashion to these works. Our goal is to distill important knowledge about how people describe fabrics, show applications like automatic captioning and retrieval, and suggest a generalization of our methodology to a larger set of material classes.

\section{Our text2fabric Dataset}
\label{sec:dataset-creation}

This section describes how we designed and created our large-scale text2fabric dataset. It consists of 45,000 rendered images depicting samples of 3,000 different fabrics, 
together with \nbvaliddescription associated descriptions in natural language. 
The full dataset, with a web-based browser to explore it, can be downloaded from the project website: \url{https://valentin.deschaintre.fr/text2fabric}.

\begin{figure*}[t]
 \centering
 \includegraphics[width=\textwidth]{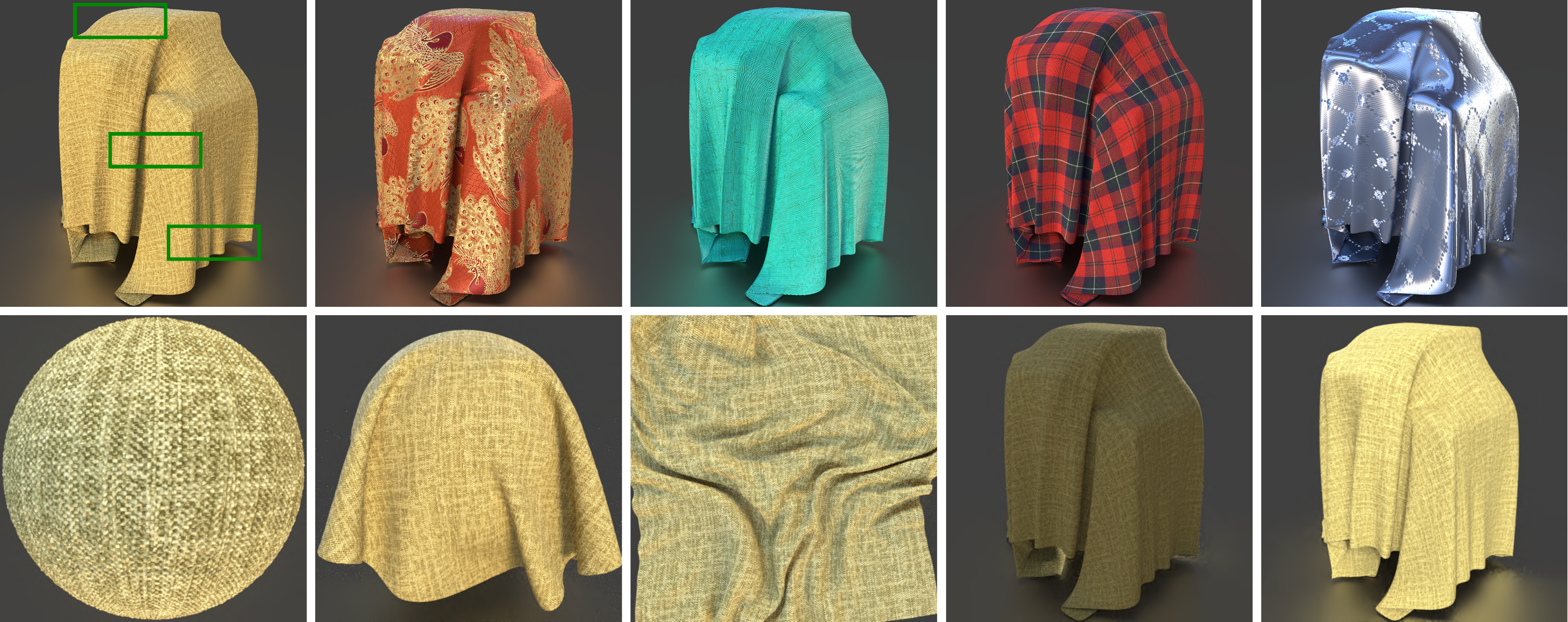}
 \caption{Representative images from our text2fabric dataset. \emph{Top row:} Five sample fabric materials, rendered with our \emph{baseline} geometry and illumination. Highlighted areas in the first image mark zoomed-in regions shown to describers. \emph{Bottom row:} Dataset images featuring three of our additional geometries and the two additional illuminations, all with the same fabric material (top row, leftmost material); from left to right: \emph{sphere}, \emph{sphere-draped} and \emph{plane-draped} geometries, and \emph{outdoor} and \emph{studio} illuminations.}
 \label{fig:dataset:images}
 \end{figure*}

\subsection{Rendered Images}
\label{subsec:dataset:images}

The 45,000 images of our dataset are generated using the Substance Stager renderer, and span 3,000 different fabric materials with a wide range of appearance. 
\new{These fabric models can be found on the Substance 3D website\footnote{~\url{https://substance3d.adobe.com/assets/allassets?assetType=substanceMaterial\&category=Fabric}}, and consist of both procedural materials generated by artists, as well as high-quality scans. Procedural materials come in the form of directed acyclic graphs, made of nodes of three different types: generators, which typically define the global bidimensional structure of the material (e.g., tiles); filters, which alter their input (e.g., colorization); and data stores, which point to external resources (e.g., raster content). Once executed by a material graph engine, these procedural models provide the material channels in the form of 2D maps, at a chosen resolution. A few parameters of the nodes of a graph are exposed as hyperparameters, so that changing them yields meaningful variations of the so-defined material. In the case of fabrics, these hyperparameters are carefully chosen so that the bounded variation they produce maps realistically to patterns and textile types encountered in the fabric industry.
In addition, each procedural material originally includes a title and some tags describing its main appearance (e.g., sportswear, upholstery, mesh). We choose not to rely on these, as they describe what the artist wanted to represent rather than how people perceive it, do not follow any particular convention, and may introduce bias in the descriptions. 
Nonetheless, this information may be used to complement our collected descriptions.
}

\new{For our task,} we first rendered images of all 3,000 different materials at 4K resolution on a \emph{baseline} geometry, carefully chosen to faithfully convey the appearance of the fabric: it contains both draped and flat areas, covering a wide range of orientations. We then selected a \emph{baseline} indoor illumination, featuring soft lighting through multiple windows.
A representative sample of the resulting fabrics can be seen in Figure~\ref{fig:dataset:images} (top row).

Additionally, to ensure robustness and help future applications (see Section~\ref{sec:applications}), we rendered the same materials using four other geometries---a sphere and a plane, in both draped and non-draped versions---, and two other illuminations---\emph{outdoor}, with direct outdoor illumination, and \emph{studio}, with strong studio indoor lighting, yielding 42,000 more images. Figure~\ref{fig:dataset:images} (bottom row) shows some examples.

Different from other existing general-purpose datasets like ImageNet or LAION\footnote{~LAION~\cite{laion400m} is the large-scale dataset on which the vision-language model CLIP~\cite{CLIP} is trained.}, our fabrics dataset constitutes a quite specific subset of images. This is illustrated, e.g., by the GLCM entropy~\cite{Haralick1973}, a measure of randomness of the images, as shown in Figure~\ref{fig:dataset:entropy}; our data yields a narrower histogram (narrower range of entropy), compared to the same number of randomly selected images from LAION \fillin{(other image statistics can be found in the supplemental material)}. The specific characteristics of our image data are relevant to its use in learning-based models, such as the ones employed in Section~\ref{sec:applications}.

\subsection{Natural Language Descriptions}
\label{subsec:dataset:text}

In the garment manufacturing industry, the description of fabrics involves specialized concepts and words such as \textit{permeability} (how much air or water it allows through), \textit{absorbency} (the ability of a fabric to take in moisture), or \textit{colorfastness} (the ability of a fabric to maintain its color and resist fading), to name a few \cite{glossary}. This specialized vocabulary is different from the one used by digital artists and practitioners in general. We thus gather our own text data to describe fabrics. 

We collected \nbvaliddescription valid descriptions of fabric appearance as free text using natural language, through a carefully controlled crowdsourcing framework (Section~\ref{subsubsec:annotation}), followed by a process of data verification and auditing (Section~\ref{subsubsec:verification}). Finally, we post-processed the resulting data (Section~\ref{subsubsec:postpro}) in preparation for the analysis described in Section~\ref{sec:dataset-analysis}.

\subsubsection{Annotation Procedure and Participants}
\label{subsubsec:annotation}
We conducted a crowdsourced user study in which participants (which we term \emph{describers}) had to provide free-text descriptions for our high-quality fabric renderings. Specifically, describers were shown one image at a time, along with three zoomed-in areas (highlighted in green in the top-left image of Figure~\ref{fig:dataset:images}), and were asked to describe the appearance of the material as precisely as possible using their own words in natural language. Describers were free to use one or several sentences for the descriptions (1-3 was recommended), and word count was limited to the range $[20..100]$ words, to prevent excessively short or long descriptions. To keep the task tractable, we gathered descriptions for our \emph{baseline} set of 3,000 images of different materials. This also encouraged describers to focus on the only changing aspect between images --the material--, familiarizing themselves with the geometry and illumination. This decision is further justified by the nature of the task, the goal of our study, and the ability of the human visual system to achieve perceptual stability and extract constant properties of materials from varying viewing conditions~\cite{tsuda2020material,fleming2017material,fleming2015perception}. 
Describers were required to do a minimum of ten descriptions, and we ensured that no describer contributed more than 9\% of the whole text data. We also ensured that, for each image, we gathered at least five \emph{valid} descriptions from different participants.

Given the specifics of the task, we required that the describers be native English speakers, had normal color vision, and were familiar with fashion or design. While we are aware that this may introduce some bias in the responses, it allows to gather a rich and accurate vocabulary. 
Prior to taking part in the study, participants underwent a short training and a qualification test. The training consisted of 
a set of instructions along with example descriptions gathered from a smaller pilot study. For the qualification test, each participant had to describe ten test fabric renderings; participants offering overly simple, poor descriptions, such as ``this is a nice fabric'', were discarded. Approximately one in four did not pass this qualification test. After this process, a total of \nbvaliddescribers describers (ages 18 through 65) went on to provide descriptions for our dataset: 45\% identified as female, 12.3\% as male, none as other gender identities, and 42.7\% preferred not to reply.

\begin{figure*}
    \centering
    \begin{tabular}{cccc}
        \subfloat[]{\includegraphics[width=0.23\textwidth]{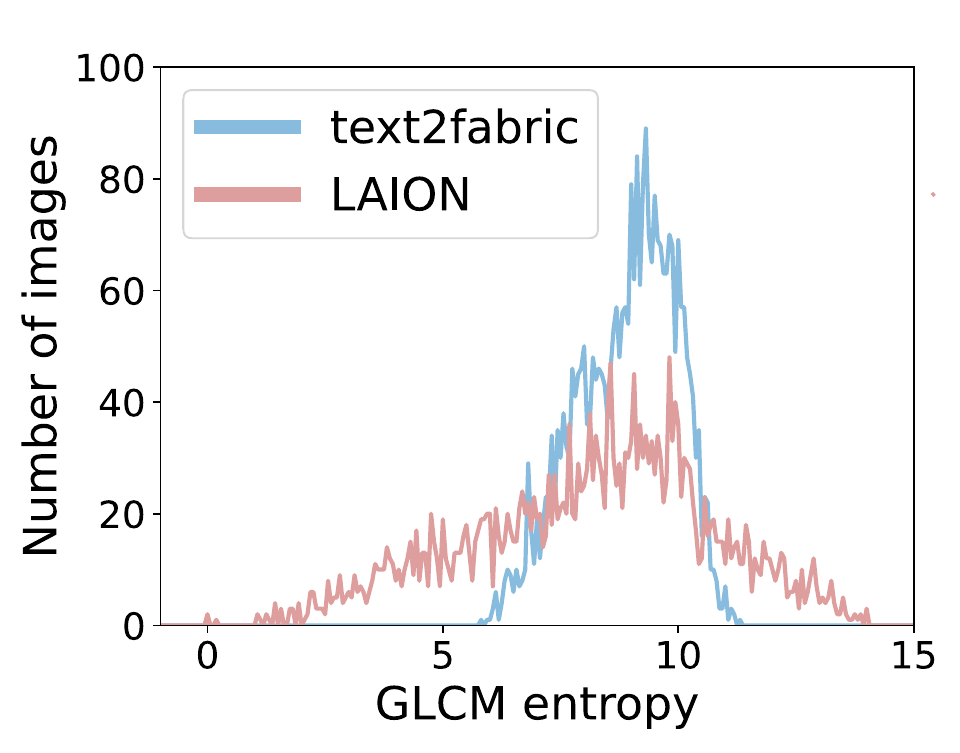}
        \label{fig:dataset:entropy}} &
        \subfloat[]{\includegraphics[width=0.23\textwidth]{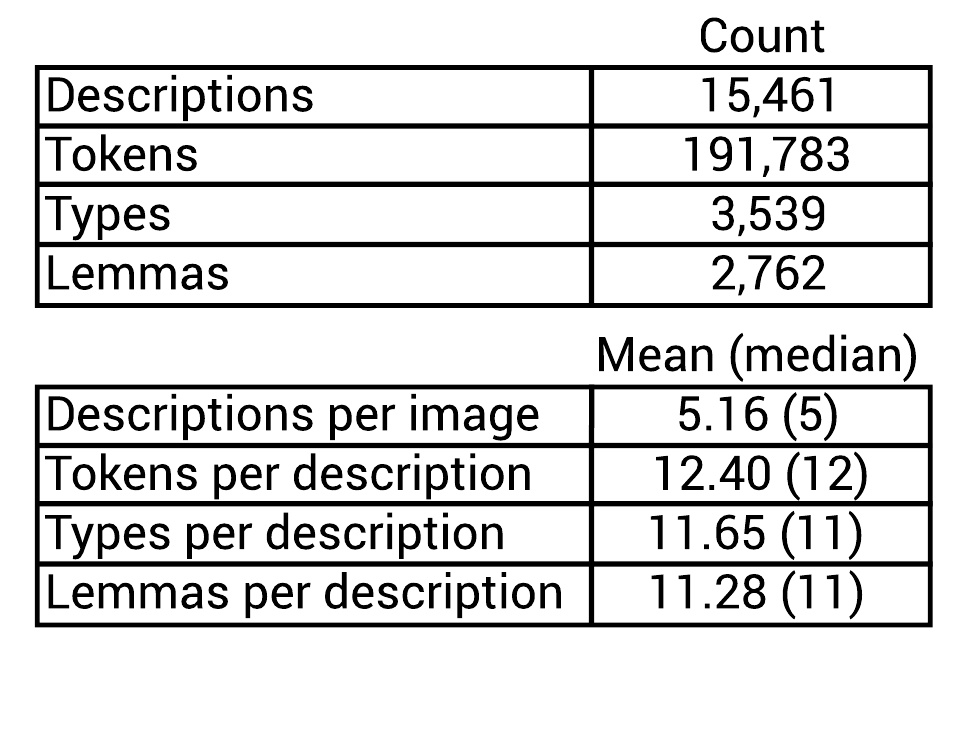}
        \label{fig:dataset:table}} &        
        \subfloat[]{\includegraphics[width=0.23\textwidth]{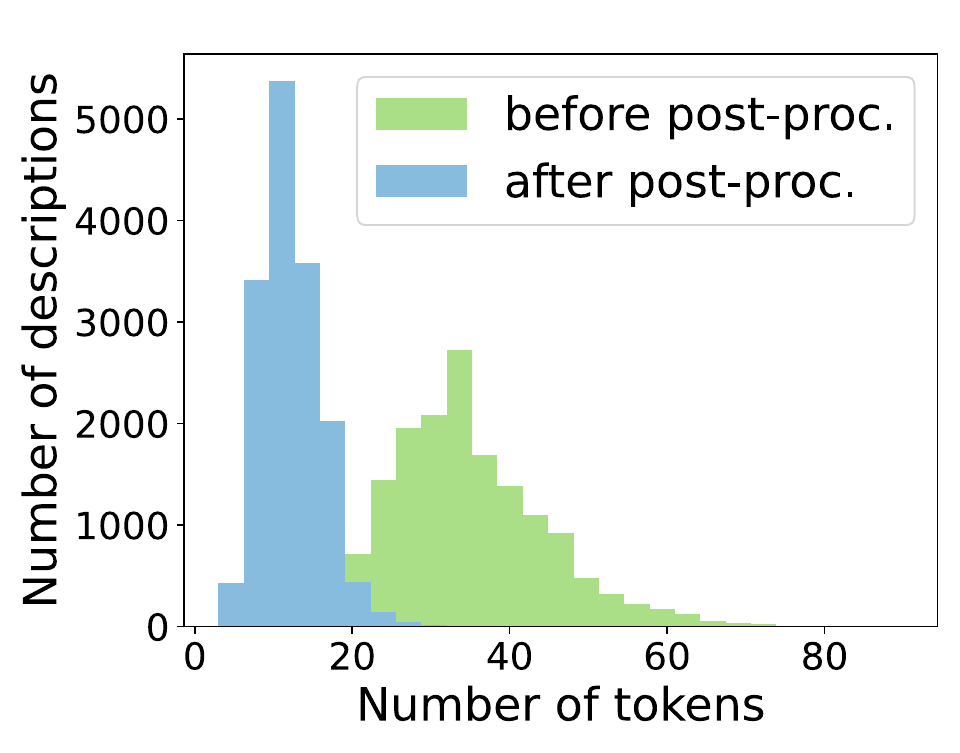}
        \label{fig:dataset:length_words}}  &
         \subfloat[]{\includegraphics[width=0.23\textwidth]{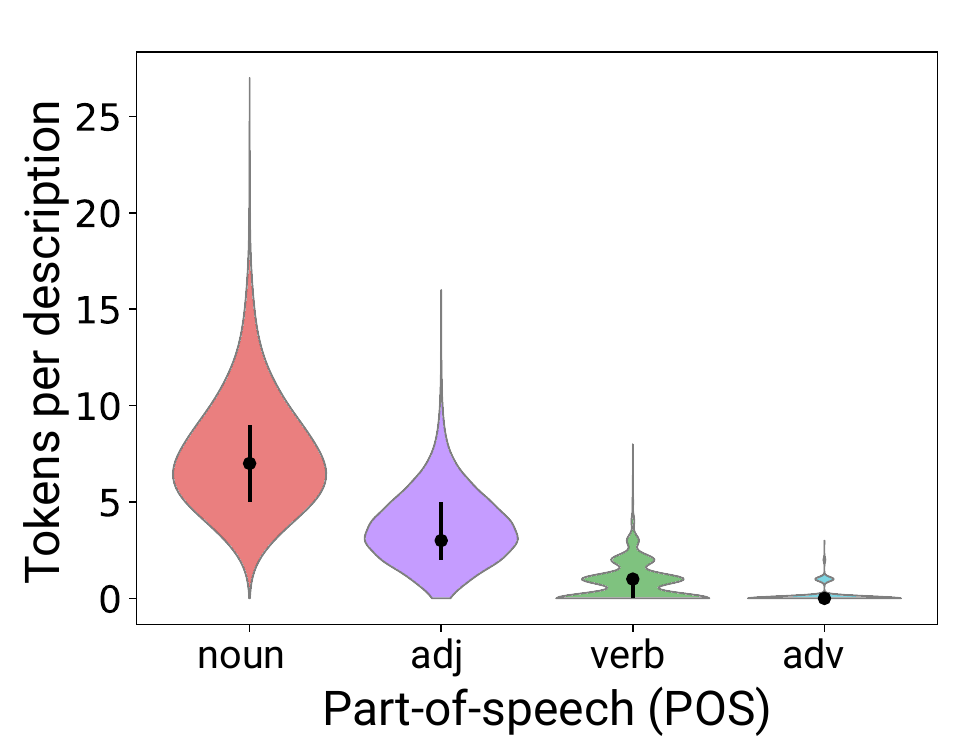}
        \label{fig:dataset:pos_tagging}}
    \end{tabular}
    \caption{
    Statistics of our text2fabric dataset. (a) GLCM entropy distribution of our image data, and of a randomly-selected subset of the LAION dataset~\cite{laion400m} of the same size as ours, for comparison. (b) Statistics of our textual data. (c) Histogram of the length, in tokens, of our descriptions before and after post-processing the text. (d) Violin plots showing part-of-speech (POS) tagging for our descriptions (black lines show the IQR, and the median is indicated by a black dot). 
    }
    \label{fig:dataset:basics}
\end{figure*}

\subsubsection{Data Verification}
\label{subsubsec:verification}

We gathered a total of 19,167 descriptions from the \nbvaliddescribers qualified describers. However, ensuring quality in free text description is a difficult task. We therefore established an additional continuous data verification protocol in which we manually audited descriptions. 
For each description, we first labeled them manually as one of four options: \emph{accepted}, or rejected due to the description being \emph{too generic}, being \emph{wrong}, or using \emph{poor grammar} to the point of hindering understandability. In addition, we also rated each description using a 5-point scale (1=totally unacceptable, 2=unacceptable, 3=acceptable, 4=very good, 5=excellent). 

Manual auditing of the full set of almost 20,000 descriptions is \new{an arduous} task. However, we found that the quality of the descriptions was highly \new{dependent} on the describer, and data quality (as given by the ratings) was fairly uniform within a describer. Therefore, auditing a randomly-selected subset of the descriptions of a participant provided a good estimate for the rest of their descriptions; for example, for participants with a rejection rate $>35\%$, we rejected all their remaining, non-audited descriptions. \fillin{More details of this process can be found in the supplemental material.} 
The data gathering process was iterative, to ensure that we had at least five descriptions for each fabric. After this process we ended up with \nbvaliddescription valid descriptions and \nbinvaliddescription invalid ones (a $19.3\%$ rejection rate). 

\subsubsection{Post-processing} 
\label{subsubsec:postpro}

Following standard natural language processing techniques, we post-process our text data by removing non-alphabetic characters, applying a spell checker, and filtering stop words. Moreover, to carry out a proper analysis we extract tokens, types, and lemmas from the descriptions~\cite{Brezina2018}. 
A \textit{token} is each occurrence of a word in a text, while a \textit{type} is each unique occurrence of a word in a text. A \textit{lexeme} corresponds to the set of alternating forms from a common root word (such as ``colors'', ``colored'' or ``coloring''), while a \textit{lemma} refers to the particular form chosen to represent a lexeme (such as ``color'' in our previous example). 
\fillin{Further details, including the spell checker and the lemmatizer we use, can be found in the supplemental material.}

The statistics of our textual data after this post-processing can be found in Figure~\ref{fig:dataset:table}. When looking at values per description, we can see that the mean and median are close, indicating that the distributions are not too skewed; we show this distribution for the case of tokens in Figure~\ref{fig:dataset:length_words}, both before and after post-processing. The table in Figure~\ref{fig:dataset:table} (bottom) further shows that the difference between the number of tokens, types and lemmas per description is not large, suggesting that our descriptions are diverse in the sense that there are not many repeated words in them.
Finally, we also perform part-of-speech tagging and classify tokens into nouns, adjectives, verbs, etc. (see Figure~\ref{fig:dataset:pos_tagging}). Our data contains mainly nouns and adjectives, as expected in texts of a descriptive nature.

\section{Understanding the visual language of fabrics}
\label{sec:dataset-analysis}

 \begin{figure}
 \centering
 \includegraphics[width=\columnwidth]{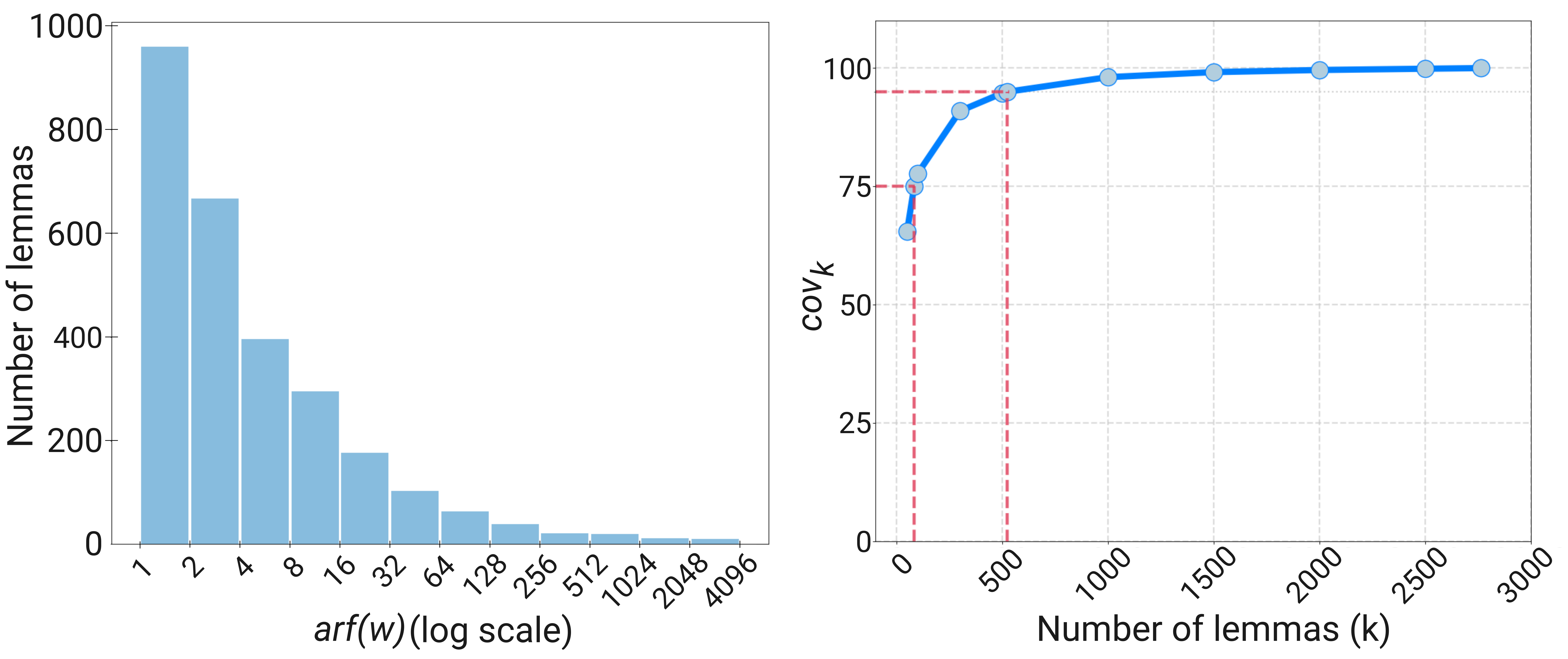}
 \caption{
 Lexicon of fabric descriptions. 
 \emph{Left:} Histogram of average reduced frequency ($\arf(\word)$) of lemmas found in our descriptions; note that the x-axis is log-scale. 
 \emph{Right:} Mean coverage of descriptions ($\coverage_k$) for different levels of $k$: a description coverage of $75\%$ is achieved with the most prominent 84 lemmas, and up to $95\%$ with the most prominent 524 lemmas.
 }
\label{fig:dataset:frequency}
 \end{figure}

We conduct a comprehensive analysis of our dataset to understand how people describe fabrics, and explore relevant questions about its characteristics. From these questions, around which this section is structured, we gather insights which help the design of tools to describe, retrieve, classify, label or edit fabrics, among others.

  \begin{figure*}
 \centering
 \includegraphics[width=0.9\textwidth]{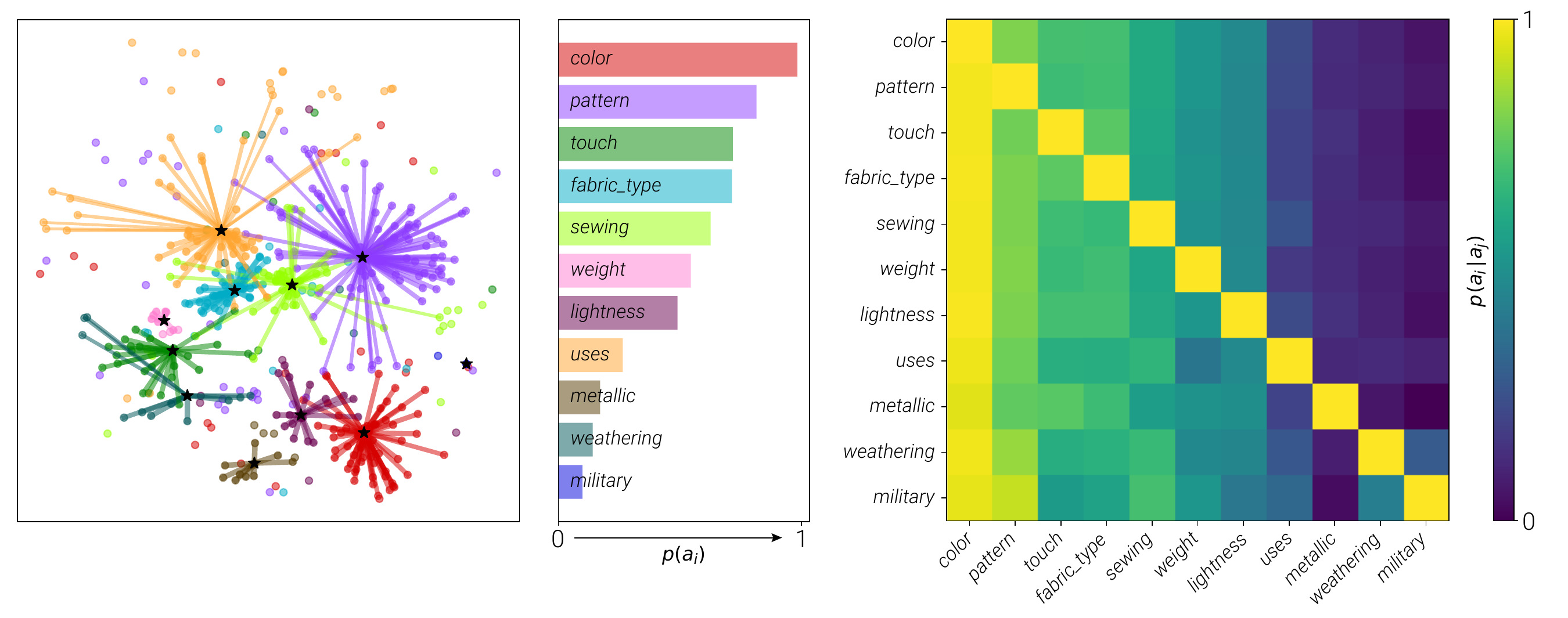}
 \caption{
 Attributes present in fabric descriptions. 
 \emph{Left:} Visualization of the lemma embeddings space and its clustering into attributes. We show all lemmas from our lexicon (524) as points in 2D space using t-SNE dimensionality reduction (300D to 2D); the color of every point indicates its associated attribute; we show a line from every non-outlier point to its attribute centroid (marked with a black star).
 \emph{Center:} Probability of occurrence of each attribute $p(a_i)$, i.e., probability of having at least one 
 occurrence of a lemma belonging to attribute $a_i$ in a description. 
 \emph{Right:} Matrix displaying $p(a_i|a_j)$, with $a_i$ in the columns and $a_j$ in the rows. Note that it is not symmetric because $p(a_i|a_j) \neq p(a_j|a_i)$. We observe how the probability of an attribute appearing is largely independent of the occurrence of other attributes.
 }
 \label{fig:dataset:attributes}
 \end{figure*}

\subsection{Is there a common lexicon when describing fabrics?}
\label{subsec:data-analysis-vocabulary}
The existence of a common vocabulary when describing fabrics is a necessary condition for any text-to-fabric application to be practical and successful. Ideally, we would like to identify a reduced set of lemmas or root words (see Section~\ref{subsubsec:postpro}) which would be sufficient for the majority of fabric descriptions.

We begin by computing the \emph{absolute frequency per lemma} $\absfreq(\word)$ in the full corpus of descriptions, where, for each lemma $\word$, the count includes occurrences of all the single words or types belonging to it. 
Prominence of lemmas, however, is not only determined by their absolute frequency, but also by their distribution; for instance, if a describer uses words belonging to a lemma often, but other describers do not, the lemma may not be prominent. We therefore complement absolute frequency with a measure of dispersion, indicating how evenly the occurrences of the lemma are distributed within the corpus. We measure this with the \textit{average reduced frequency} ($\arf(\word)$)~\cite{Savicky2002,Brezina2018}, which modulates $\absfreq(\word)$ with the dispersion of $\word$ (\fillin{details on the computation can be found in the supplemental material}). The histogram of $\arf(\word)$ (Figure~\ref{fig:dataset:frequency} (left)) 
shows how over one third of the lemmas have an $\arf$ value below $2$, meaning that they are seldom used, or used by a single describer. This confirms the intuition that a reduced subset of lemmas should suffice for fabric description.

We next examine how small this reduced lexicon can be. 
We first use $\arf(\word)$ as ranking criterion to find the subset $\mathcal{W}_k$ of the most prominent $k$ lemmas, for increasing values of $k \in [1..N_\word]$, with $N_\word=\nblemmaspostpro$ the number of lemmas in our corpus. 
For each $\mathcal{W}_k$, we then compute the \textit{coverage} of a description $d$ as
\begin{equation}
\nonumber
\coverage_k(d) = \frac{n_k(d)}{n_{tot}(d)},
\end{equation}
where $n_k$ is the number of lemmas from subset $\mathcal{W}_k$ present in description $d$, and $n_{tot}(d)$ is the total number of lemmas of such description. 
As the plot in Figure~\ref{fig:dataset:frequency} (right) shows, a common lexicon of 84 lemmas is enough to cover 75\% of the fabric descriptions, while to cover 95\% we only need 524 lemmas, which we define as our fabric-specific lexicon \new{(examples of these lemmas can be found in the supplemental material).}

\subsection{Are there key attributes in the descriptions?}
\label{subsec:data-analysis-attributes}

Finding common attributes is useful to understand how we internally represent and think about fabrics. From our reduced 524-lemma lexicon, we seek now to identify common attributes that these lemmas may relate to. We approach this as a clustering problem, and develop a methodology based on affinity propagation and similarity between lemmas. In particular, we leverage embeddings of the lemmas 
provided by \textit{ConceptNet Numberbatch}~\cite{speer2017conceptnet}, which combines both distributional semantics and relational knowledge\footnote{~We use the implementation from \url{https://github.com/commonsense/conceptnet-numberbatch}. We refer the reader to the original ConceptNet paper~\cite{liu2004conceptnet}, as well the ConceptNet Numberbatch extension~\cite{speer2017conceptnet} for more details.}.
This leads to the identification of the main attributes that people focus on when describing fabrics, as well as a distribution of our lexicon into those attributes (\fillin{we provide a description of this process in the supplemental material}). 
This results in eleven key attributes describing fabrics: \emph{color}, \emph{lightness}, \emph{metallic}, \emph{pattern}, \emph{fabric\_type}, \emph{sewing}, \emph{touch}, \emph{weight}, \emph{use}, \emph{weathering}, and \emph{military}\footnote{~This last attribute reflects the significant amount of samples of a military nature in our dataset and may not generalize to others, see also Section~\ref{sec:discussion}.}, 
and are shown in Figure~\ref{fig:dataset:attributes} (left) using t-SNE dimensionality reduction~\cite{tsne}.

In Figure~\ref{fig:dataset:attributes} (center), we show the probability of occurrence $p(a_i), i \in [1..\nattr]$ of each attribute $a_i$, where $\nattr=11$ is the number of attributes. It expresses the probability that 
there is at least one occurrence of a lemma belonging to the attribute  
in any given description. This illustrates the relative importance of each attribute: for instance, it reveals that \emph{color}, \emph{pattern}, \emph{touch} and \emph{fabric\_type} are present in more than 70\% of the descriptions.

Moreover, we look into whether certain attributes tend to appear together in the descriptions; to that end we compute 
$p(a_i|a_j)$, $i,j \in [1..\nattr], i \neq j $, i.e., the probability of attribute $a_i$ being present in a description that contains attribute $a_j$. 
Figure~\ref{fig:dataset:attributes} (right) plots these probabilities for all attributes (note that the resulting matrix is non-symmetric, since $p(a_i|a_j) \neq p(a_j|a_i)$). We observe that, in general, the presence of a given attribute in a description is not heavily dependent on the presence of any other attribute. This is indicated by the relatively uniform values along each column,  
and is a result of the large variety of appearances present in our dataset, exhibiting many different combinations of attributes.

\subsection{Do descriptions follow a common structure?}
\label{subsec:data-analysis-structure}

We next look at the structure of descriptions by analyzing the order of appearance of the different attributes. 
Specifically, we compute a \emph{rank product} for each attribute as 
\begin{equation}
\nonumber
\Psi(a) = (\prod_{i=1}^{\ndescr}r_{a,i})^{1/\ndescr},
\end{equation}
where $r_{a,i}$ is the \emph{rank} of attribute $a$ in description $d_i, i \in [1..\ndescr]$~\cite{Rubinstein_SA2010}. The rank is given by the first appearance of a lemma belonging to an attribute in a description; thus, lower rank products indicate that the attribute tends to appear earlier in the descriptions. 

Table~\ref{tab:dataset:rank-product} shows the resulting ranking of attributes. To evaluate whether the differences in ordering are significant, we perform a Kruskal-Wallis test (a non-parametric extension of ANOVA, since rankings are an ordinal value and typically cannot be assumed to follow a normal distribution), which shows that there is a significant difference between attributes ($H(10)=8235.53$, $p<.0001$). 
A subsequent pairwise comparisons test allows us to identify groups of attributes where  there is no significant difference between their mean ranks (also shown in Table~\ref{tab:dataset:rank-product}).
\fillin{The rank histograms per attribute can be found in the supplemental material}.

\begin{table*}
\caption{Attributes sorted by rank product, indicative of their order of appearance within a description. Lower rank products indicate that the attribute tends to appear earlier in the descriptions. Attributes grouped together in the table yield no significant difference between their mean ranks.} 
\begin{tabular}{|l|c|c|ccc|cc|c|c|c|c|}
\hline
Attribute $a$  & color & lightness & sewing & metallic & pattern & weight & military & fabric\_type & weathering & touch & use \\ 
Rank product $\Psi(a)$ & 2.25  & 2.39      & 2.75   & 2.77     & 2.86    & 2.95   & 3.06     & 3.17         & 3.46   & 3.73  & 4.25   \\ \hline
\end{tabular}
 \label{tab:dataset:rank-product}
\end{table*}

\subsection{Does the same fabric elicit similar descriptions?}
\label{subsec:dataset-analysis-similarity}

We measure similarity between descriptions using two state-of-the-art NLP models that have been shown to work well on Semantic Textual Similarity (STS): sentence-T5~\cite{ni2021sentenceT5}, designed to provide sentence embeddings from text-to-text  transformers,
and MPNet~\cite{song2020mpnet}, shown to work well for semantic search using sentence embeddings~\cite{reimers2019sbert}. Specifically, we compute cosine similarity between the embeddings of our full descriptions, as in the original publications.

To compute the \emph{intra-image} description similarity (similarity between descriptions of the same fabric), we average over all pairwise comparisons  
in our whole corpus, provided that the two members of a pair belong to the same image. To compute the
\emph{inter-image} description similarity (similarity between descriptions of different fabrics), we average over all pairwise comparisons in our whole corpus, provided that the two members of a pair belong to different images. 

The results, shown in Table~\ref{tab:dataset:intra-inter}, yield a high intra-image similarity (cosine similarities are bounded between -1 and 1), suggesting that the same fabric does indeed elicit similar descriptions by different people. Compared to the inter-image similarity (which one may treat as a baseline), the average intra-image similarity is larger for both models. 
To test whether these differences are  statistically significant, we resort to ANOSIM (analysis of similarities)~\cite{clarke1993anosim, warton2012anosim}. ANOSIM works on all pairwise similarities (or distances) between points (descriptions), and is designed to test the null hypothesis that the similarity between groups (inter-image) is greater than or equal to the similarity within groups (intra-image). We use a p-value of $0.05$ to indicate significance
\new{and the test statistic as the measure of effect size~\cite{somerfield2021analysis}. This value is constrained to $[-1,1]$, with $1$ indicating very high intra-image similarity with respect to inter-image similarity, and negative values indicating higher inter-image similarity. Results of this analysis show reasonably high intra-image similarity with respect to inter-image similarity, 
confirming that the difference is statistically significant (see Table~\ref{tab:dataset:intra-inter}).} 

\begin{table}
\caption{Similarity between descriptions of the same image (intra-image) and descriptions of different images (inter-image). We report, for two state-of-the-art sentence embeddings (sentence-T5 and MPNet): average intra-image and inter-image similarities (and associated standard deviations), test statistics for ANOSIM, and associated p-value (we use a p-value of 0.05 to indicate significance). The descriptions in our dataset exhibit high intra-image similarity, and the statistical test shows that intra-image similarities are significantly larger than the inter-image ones.
 }
\small
\begin{tabular}{rcccc}
\cline{2-3}
\multicolumn{1}{l|}{}             & \multicolumn{2}{c|}{Similarity}                                              & \multicolumn{1}{l}{}                & \multicolumn{1}{l}{} \\ \cline{2-5} 
\multicolumn{1}{c|}{}             & \multicolumn{1}{c|}{Intra-image}     & \multicolumn{1}{c|}{Inter-image}     & \multicolumn{1}{c|}{Test Statistic} & \multicolumn{1}{c|}{p-value}              \\ \hline
\multicolumn{1}{|r|}{sentence-T5} & \multicolumn{1}{c|}{0.874 (0.037)} & \multicolumn{1}{c|}{0.822 (0.037)} & \multicolumn{1}{c|}{0.694}         & \multicolumn{1}{c|}{0.001}                 \\ 
\multicolumn{1}{|r|}{MPNet}    & \multicolumn{1}{c|}{0.704 (0.087)} & \multicolumn{1}{c|}{0.627 (0.083)} & \multicolumn{1}{c|}{0.497}         & \multicolumn{1}{c|}{0.001}               \\ \hline
\end{tabular}
\label{tab:dataset:intra-inter}
\end{table}

\section{Large Vision-Language Model Coupling}
\label{sec:applications}

Our dataset links the appearance of fabrics with natural language, helping to better understand how people describe such materials despite their semantic proximity. 
Besides, it provides high quality image and associated text data, in large albeit lower amounts than those present in very large-scale datasets used to train recent, very successful vision-language models (see Section~\ref{subsec:apps:clip_blip}). 
In this section, 
we explore applications of our dataset with such models, and to what extent a relatively low amount of high-quality, specialized data improves over their native versions for specific areas such as material appearance. Specifically, we demonstrate text-based fine-grained retrieval, image-based search, and description generation, as well as an improvement of invariance of the image latent representations to light and geometry changes, contributing to, e.g., a more robust notion of appearance similarity. \fillin{While we show here varied results and evaluations, please also refer to the supplemental material for additional examples.}

\subsection{Large Vision-Language Models}
\label{subsec:apps:clip_blip}

Recent progress in joint text and image encoding has been enabled by large vision-language models. In this section in particular, we fine-tune and compare to two of the most widely-used models: CLIP~\cite{CLIP} and BLIP~\cite{BLIP}. 

CLIP is a neural model composed of two encoders, one for each modality (text and image), which are trained using pairs of text and images. The method relies on contrastive learning~\cite{chen2020simple} to encourage encodings of texts and images to lie close to one another in latent space. This has been shown to draw very interesting connections based on the data it is trained on~\cite{goh2021multimodal}. CLIP is particularly powerful thanks to the vast LAION dataset on which it is trained, containing 400 million image-text pairs gathered from the internet. Different encoder architectures have been published, but in this paper we use the ViT-B/16 version, which relies on a visual transformer~\cite{dosovitskiy2020image} with a patch size of $16\times16$.

BLIP is a combination of networks trained jointly, including a pair of encoders, similar to CLIP. Moreover, BLIP also contains a generative head, trained jointly with the rest of the network, enabling it to generate captions corresponding to an image. It is also trained on hundreds of millions of images, including a self-supervised augmentation mechanism called ``CapFilt''. Similarly to CLIP, we use the ViT-B version of the network. 

While both CLIP and BLIP are trained on very large-scale datasets, the text data to which they are exposed is limited to low quality online captions of images. We will show, in the remainder of this section, that a small amount of high quality data is sufficient to significantly improve the networks' sensitivities to specialised concepts.
\new{In the following experiments, we use the models published by SalesForce and OpenAI.
}

\subsection{Text-Based Fine-Grained Retrieval}
\label{subsec:apps:retrieval}

Given a query in the form of a text description, the goal of this first application is to retrieve fabric samples that match the query. 
Improving search performance in large datasets is increasingly important as the number and size of libraries and datasets increase~\cite{Substance3DAssets,megascan}.
Different from generic text-based retrieval, which aims at finding images of objects in different classes such as ``chairs'', ``cars'', or ``people'', we target the more difficult case of \textit{fine-grained} retrieval~\cite{qi2021Retrieval}, i.e., finding the right instance despite significant semantic similarities within the dataset.

\subsubsection{Implementation}
\label{subsubsec:implementation-clip}
We fine-tune CLIP~\cite{CLIP}, 
starting from the VIT-B/16 pre-trained model (we term this pre-trained model \emph{native} CLIP). Our dataset is split in 12,334 training and 3,129 test descriptions, ensuring that no procedural variation of a given material in the training data is used for testing. 
During training we split the descriptions by sentence,  
resulting in 45,871 (36,565 train/9,306 test) individual sentence descriptions for 3,000 (2,393/607) materials.
We train the network for 12 epochs \new{(at which point the performance levels off, with small gains until epoch 19)}, using renderings of the training materials on four geometries (\emph{baseline}, \emph{sphere}, \emph{sphere\_draped}, \emph{plane}) and a single illumination (\emph{baseline}). We use a learning rate of $1e^{-6}$ with a linear schedule with $200$ warm-up steps for the Adam optimizer, with $\beta_1 = 0.9$, $\beta_2 = 0.99$ and batch size $128$. This takes five hours to train on a single Nvidia RTX3090 GPU. \new{For inference, execution time is 0.46 seconds for a batch of 64 images.}

\subsubsection{Results}
Since we have ground-truth data (image-description pairs in our test dataset), we can evaluate retrieval of the \emph{correct} material given an input description. Additionally, for any given generic query, the retrieval application should return relevant results. 
The search operation presented in this section is made over the entire 3,000 materials on our \emph{baseline} geometry and illumination unless specified otherwise; in all cases, neither the descriptions used as queries nor the correct images or materials have been seen during training.

\paragraph{Quantitative analysis}
Given our test set descriptions, we evaluate retrieval in the complete material database and report the top-K recall results of our fine-tuned CLIP, with $K \in \{1,5,10,20,100\}$, in Table~\ref{tab:application:retrieval}. We also include results for \emph{native} CLIP, \emph{native} BLIP\footnote{~For details on the implementation of native BLIP please refer to Section~\ref{subsubsec:implementation-blip}.}, \new{and BLIP trained on our data only (BLIP \emph{no pretrain}).} 
Compared to native CLIP/BLIP, we achieve $4.8$/$4$ times better top-1 retrieval rate and maintain at least $2.12$ times better results for all top-K results, showing that our dataset makes CLIP more sensitive to fabric-specific concepts. This also shows that our fine-tuned model is capable of retrieving a fabric sample from its description alone, which requires strong feature discrimination in a semantically similar dataset. 
\new{The comparison to BLIP trained on our data only (no pretrain) shows that our model significantly benefits from the original model training, leveraging the priors provided by its large corpus of text.}

\begin{table}
 \caption{Top-K retrieval results on the \emph{baseline} geometry for native CLIP, native BLIP, \new{BLIP trained on our data only (BLIP no pretrain)} and our fine-tuned model.
 }
\small
\begin{tabular}{r|c|c|c|c|}
 & Native CLIP & Native BLIP & \new{BLIP no pretrain} & Ours \\ \hline
 Top-1 & 2.94\% & 3.42\% & 1.60\% & \textbf{13.81\%} \\
 Top-5 & 8.31\% & 9.94\% & 5.98\% & \textbf{33.91\%} \\
 Top-10 & 12.59\% & 14.60\% & 10.64\% & \textbf{46.76\%} \\
 Top-20 & 18.37\% & 20.17\% & 17.00\% & \textbf{59.76\%} \\
 Top-100 & 41.29\% & 34.26\% & 34.36\% & \textbf{87.63\%} \\ \hline
\end{tabular}
 \label{tab:application:retrieval}
\end{table}

\begin{table}
 \caption{Top-K retrieval results on the \emph{plane\_draped} geometry, unseen during training, for native CLIP, native BLIP, our model fine-tuned on only one geometry (\emph{baseline}), and our model (which is fine-tuned on four geometries, not including \emph{plane\_draped}). 
 }
 \small
 \begin{tabular}{r|c|c|c|c|}
 & Native CLIP & Native BLIP & Ours (1 gm.) & Ours (4 gm.) \\ \hline
 Top-1 & 1.34\% & 2.27\% & 5.98\% & \textbf{7.38\%}  \\
 Top-5 & 5.02\% & 7.03\% & 16.55\% & \textbf{22.95\%} \\
 Top-10 & 7.93\% & 10.9\% & 23.94\% & \textbf{31.77\%} \\
 Top-20 & 12.66\% & 15.05\% & 34.16\% & \textbf{43.72\%} \\
 Top-100 & 32.18\% & 29.05\% & 64.59\% & \textbf{75.93\%} \\ \hline
\end{tabular}
 \label{tab:application:retrieval_unseen}
\end{table}

\begin{figure}
 \centering
 \includegraphics[width=0.8\columnwidth]{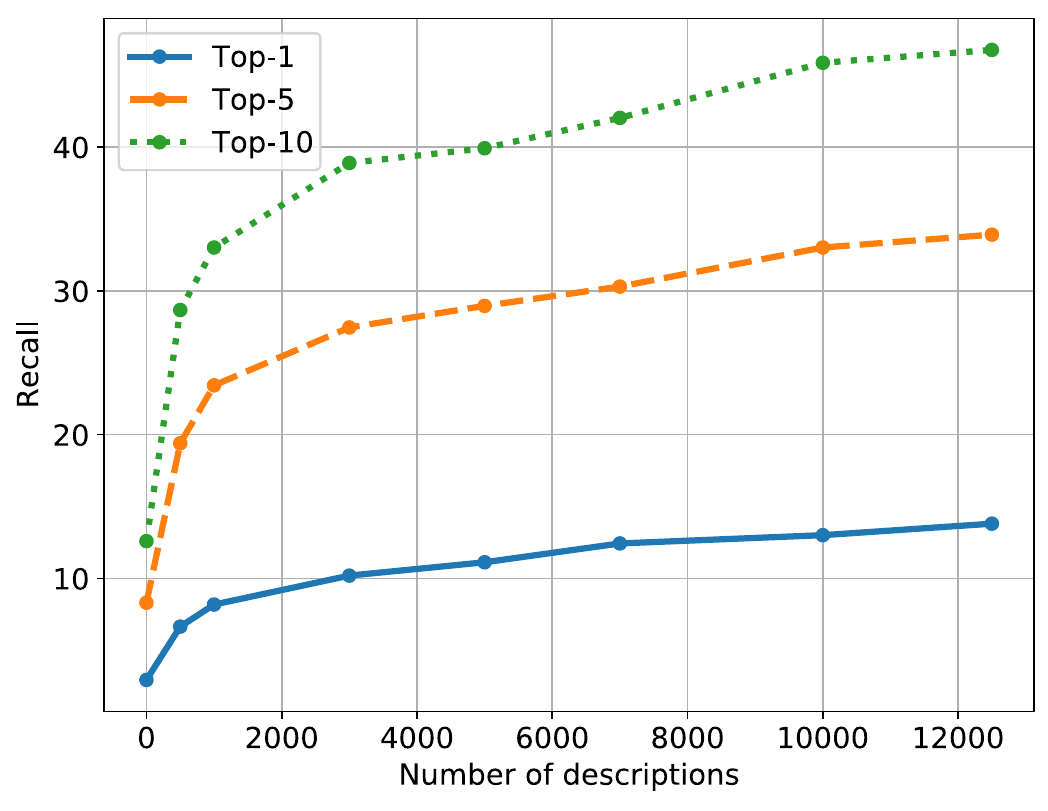}
 \caption{Evolution of text-based retrieval results (top-1, top-5 and top-10 recall performance) with the number of descriptions available for fine-tuning. Native CLIP performance corresponds to the case of zero descriptions available for fine-tuning.
 }
 \label{fig:application:retrievalAblationAdescriptionsnb}
 \end{figure}

To evaluate the ability of our model to generalize to other geometries, Table~\ref{tab:application:retrieval_unseen} reports retrieval recall results on a geometry unseen during training (\emph{plane\_draped}). 
We see that, despite the unseen geometry being challenging (all methods have lower retrieval results than with the \emph{baseline} geometry), our fine-tuned model still significantly benefits from our dataset. Furthermore, we also include a comparison to our model fine-tuned on images from just one geometry (\emph{baseline}), showing that the training on different geometries improves the generalization of the method.

\begin{figure*}
 \centering
 \includegraphics[width=\textwidth]{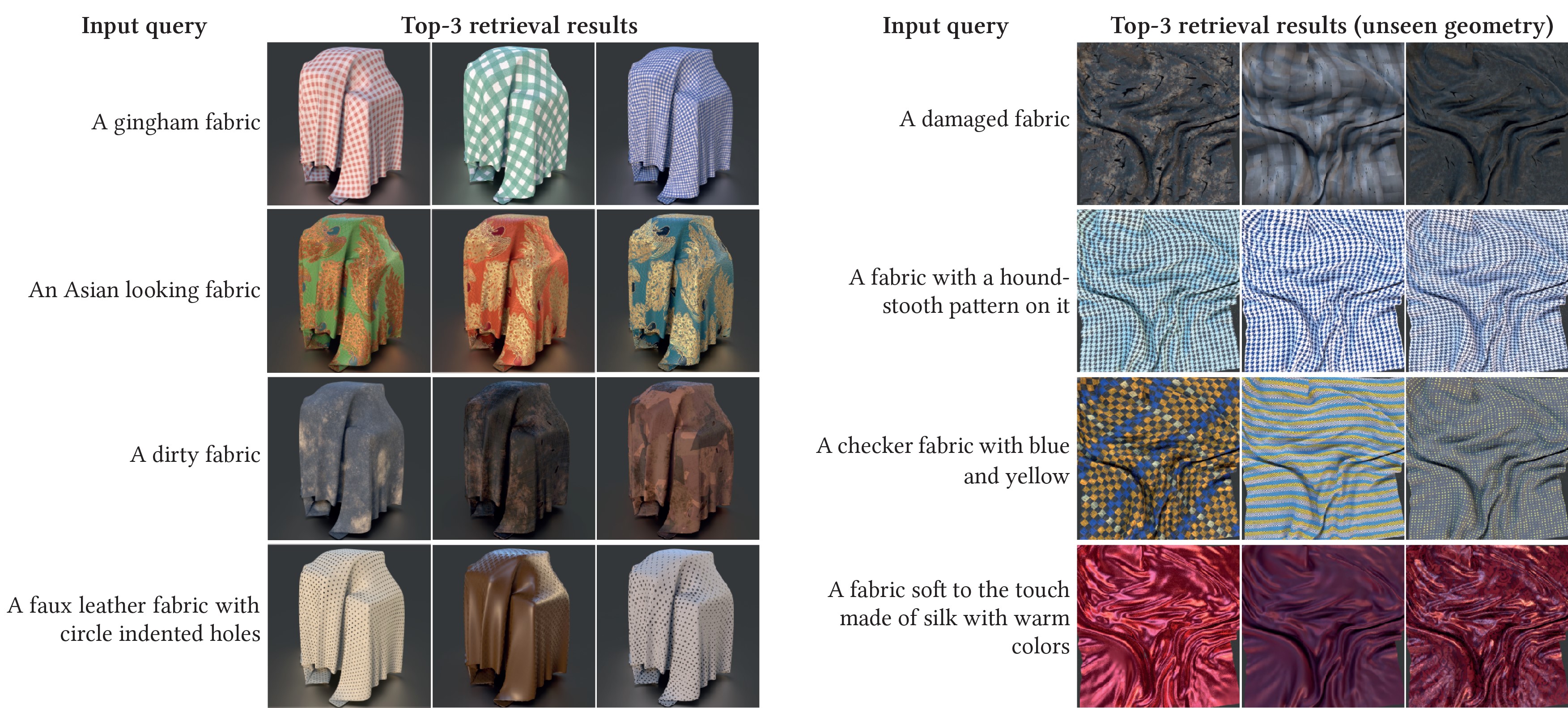}
 \caption{
 Text-based fine-grained retrieval, evaluating the sensitivity of our fine-tuned representation to varied domain-specific concepts on two different geometries. We show input text queries, and the top-3 retrieval results using our fine-tuned model. \emph{Left:} Retrieval results on the \emph{baseline} geometry, seen during fine-tuning. \emph{Right:} Retrieval results on the \emph{plane\_draped} geometry, unseen during fine-tuning. 
 Our model retrieves relevant results for aspects related to different attributes, and for both high-level and more specific queries.
 }
 \label{fig:application:retrieval}
 \end{figure*}

\begin{figure*}
 \centering
 \includegraphics[width=\textwidth]{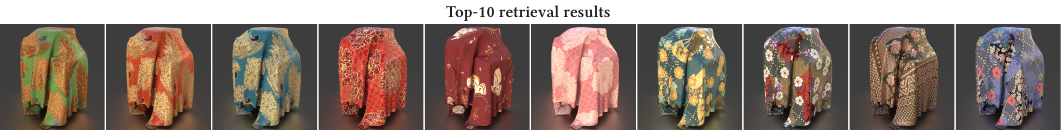}
 \caption{\new{Top-10 results of text-based fine-grained retrieval with our fine-tuned model for the query ``An Asian looking fabric''. Although the closest samples (top-3, also shown in Figure~\ref{fig:application:retrieval}) have similar appearance, we observe more diverse results (while still relevant) when increasing the number of images returned.}
 }
  \label{fig:application:top10}
 \end{figure*}

We also assess the required size of a specialized dataset such as ours for fine-tuning general purpose models. To do so, we plot the top-K retrieval results as we vary the number of descriptions used in the training in Figure~\ref{fig:application:retrievalAblationAdescriptionsnb}. We see how the model significantly benefits from the first 2,000 descriptions, and how the marginal improvement rate then starts diminishing. At constant number of descriptions, we also evaluate whether more images with \new{fewer} descriptions is preferable to \new{fewer} images with more descriptions. We find that using 1,500 images with 5 descriptions per image is equivalent in retrieval recall to using 2,500 images with 3 descriptions per image: Results in both cases are close, indicating a similar impact between image and description diversity. 
More precisely, 1,500 images with 5 descriptions each yield top-1/5/10 retrieval results of 12.56/31.38/44.1\%; in comparison, 2,500 images with 3 descriptions each yield top-1/5/10 retrieval results of 12.66/31.16/43.69\%.
\new{In addition, we also include a quantitative evaluation on negative queries in the supplemental material}.

\vspace{-0.5pt}
 
\paragraph{Qualitative analysis}
  \begin{figure*}
 \centering
 \includegraphics[width=\textwidth]{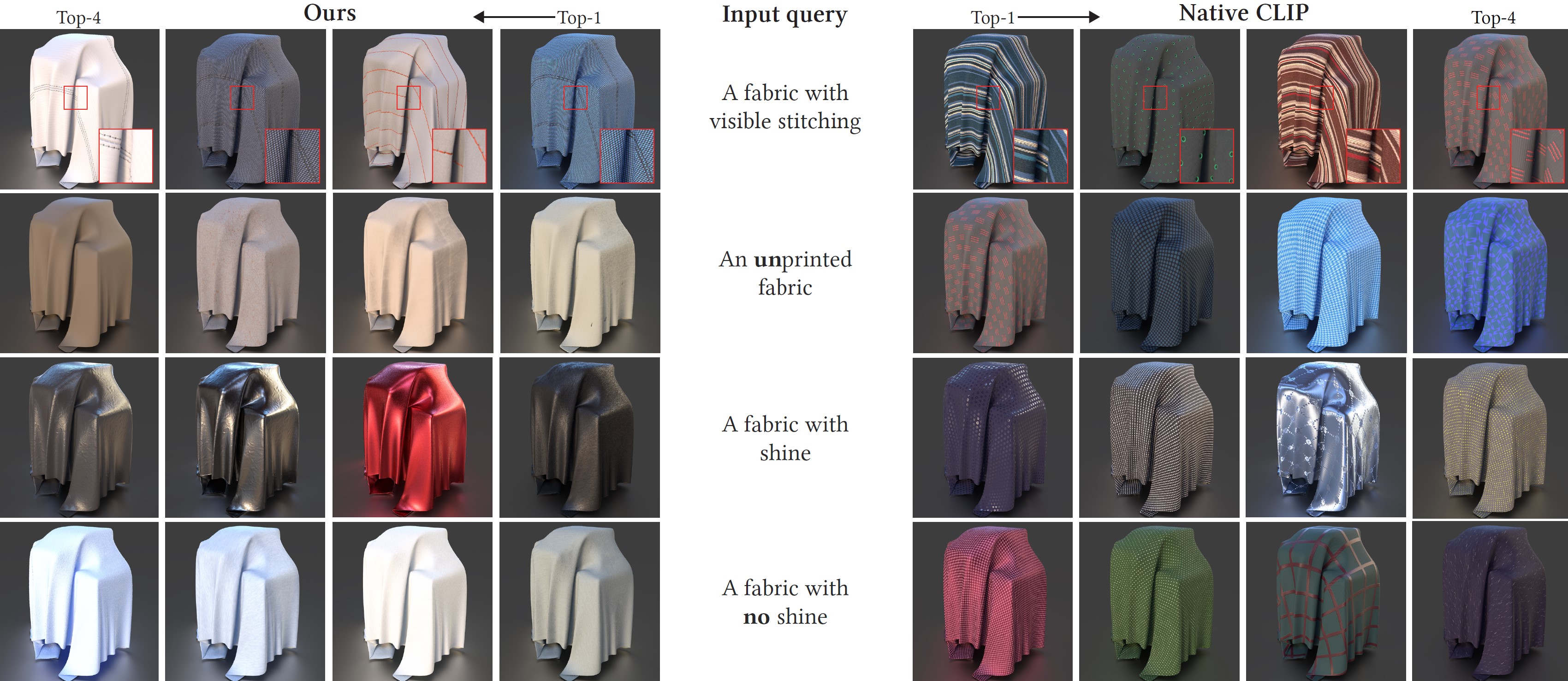}
 \caption{
 Text-based fine-grained retrieval. We evaluate the sensitivity of our fine-tuned model to different text queries, including negative ones, and compare it with native CLIP. Our model is more sensitive to domain-specific concepts and can better handle negative queries.
 }
 \label{fig:application:retrieval_sensitivity}
 \end{figure*}

 \begin{figure}
 \centering
 \includegraphics[width=1.0\columnwidth]{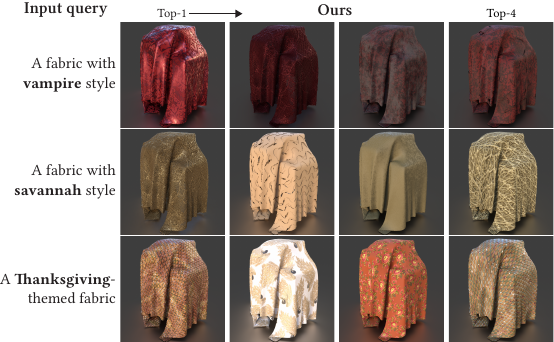}
 \caption{\new{Text-based fine-grained retrieval with out-of-distribution queries that contain concepts not included in our dataset descriptions (marked in bold). 
 We show top-4 retrieval results using our fine-tuned model.}
 }
  \label{fig:application:out_of_distrib}
 \end{figure}

To qualitatively evaluate the performance of our fine-tuned model, 
we provide results retrieved from natural language queries in Figure~\ref{fig:application:retrieval}, showing that the retrieved materials exhibit the desired properties, not only in a geometry seen during training\footnote{~Note that the geometry has been seen during training, but the materials and queries have not.} (\emph{baseline}), but also in unseen geometry (\emph{plane\_draped}). 
\new{As expected, diversity increases as we look at more returned samples. This can be seen in our ``Asian looking'' prompt; while the top-3 results in Figure~\ref{fig:application:retrieval} contain similar results due to our space being partly organized with respect to visual features, more diverse results appear when visualizing the top-10 results, as shown in Figure~\ref{fig:application:top10}.}

In Figure~\ref{fig:application:retrieval_sensitivity}, we show a more systematic evaluation with positive and negative queries, corresponding to prominent fabrics concepts extracted from the dataset (see Section~\ref{sec:dataset-analysis}), and include a comparison to native CLIP. Results confirm that our fine-tuned model is effective in the retrieval, and more sensitive to fine-grained descriptions, while native CLIP struggles with specialized concepts (e.g., stitching) and negative wording. 
Interestingly, despite the relatively small amount of data used in our fine-tuning (compared to the hundreds of millions of image-text pairs required to train CLIP and BLIP), we observe significant improvement in material retrieval for the class of interest (fabrics). Furthermore, these experiments highlight the limitations existing in the representations of Large Vision-Language Models for fine-grained appearance concepts.

\new{Finally, we evaluate the limits of modeling out-of-distribution queries, containing concepts that do not appear in our dataset descriptions. As shown in Figure~\ref{fig:application:out_of_distrib}, our model finds reasonable results for these queries (e.g., for the case of ``Thanksgiving-themed'' we obtain a variety of autumnal brown and orange fabrics), suggesting that our model preserves its broader priors without overfitting to our dataset.}

   \begin{figure*}
 \centering
 \includegraphics[width=\textwidth]{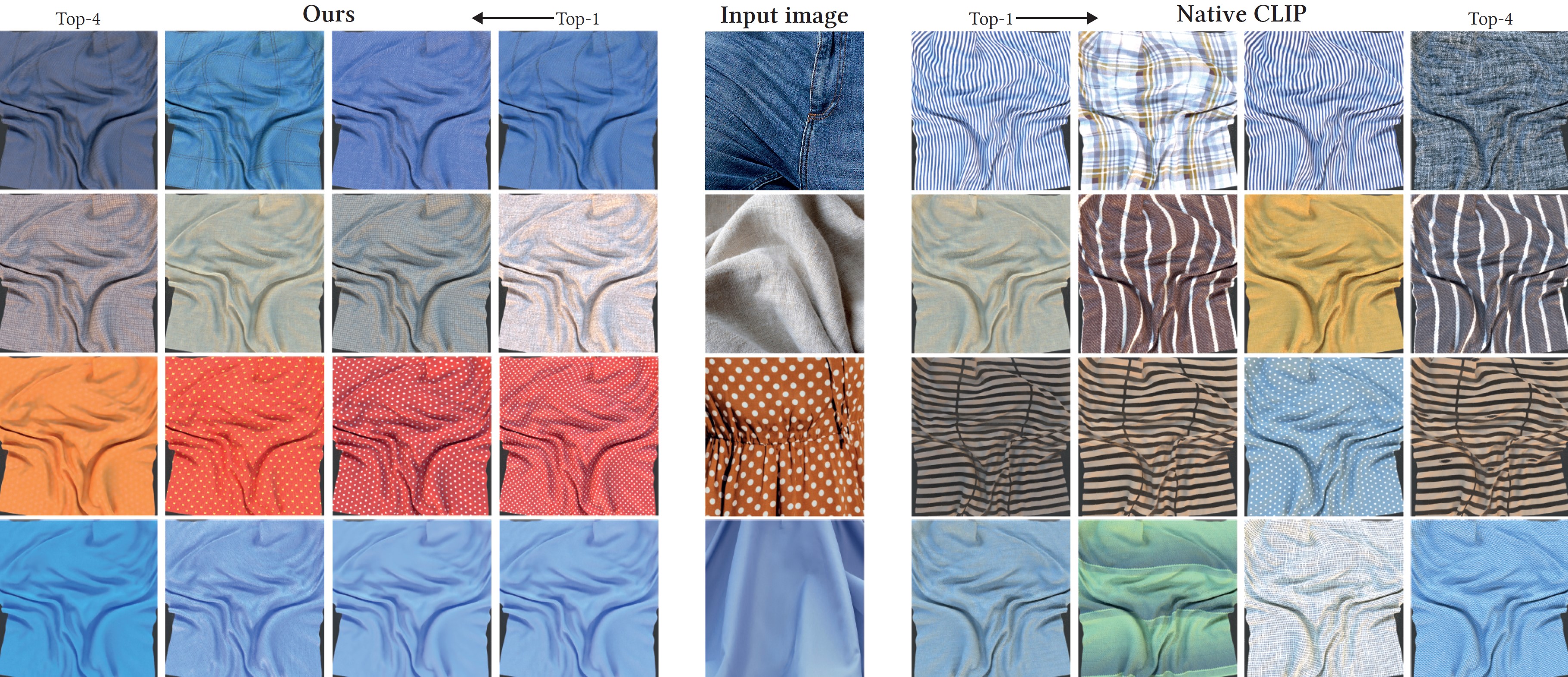}
 \caption{Image-based search with real photographs as input (\emph{middle column}), using our fine-tuned model (\emph{left}) and using native CLIP (\emph{right}). The search is performed on image data from a single geometry (\emph{plane\_draped}), unseen during the fine-tuning. 
 Our model, unlike native CLIP, is capable of retrieving results that are in close correspondence with the input, while circumventing the prominent characteristics arising from the macroscopic geometric structure. 
 }
 \label{fig:application:imageRetrieval}
 \end{figure*}

\subsection{Image-Based Search}
\label{subsec:apps:search}

We continue our evaluation by studying image-based search using real images as input. 
We do it by leveraging our fine-tuned CLIP model (see Section~\ref{subsec:apps:retrieval}), as well as native CLIP for comparison.
Specifically, we compute the normalized embedding---using either native CLIP or our fine-tuned model---of the input image, and compute its cosine distance to the normalized embeddings of the candidates from our dataset. These candidates are the 3,000 materials in our dataset, rendered on a certain geometry (or set of geometries in Section~\ref{subsec:apps:invariance}).
Figure~\ref{fig:application:imageRetrieval} shows results on the \emph{plane\_draped} geometry (unseen during training) for both our fine-tuned model and native CLIP.
We can observe that native CLIP is strongly influenced by the geometrical macrostructure present in the input image, and fails at guiding the retrieval process by the material mesostructure, patterns and reflectivity properties expressed in the input. On the contrary, our fine-tuned model succeeds at fetching results with similarities existing at material scale, bypassing the strong features stemming from the supporting 3D shape.  

\subsection{Caption Generation}
\label{subsec:apps:caption}

Caption generation aims at creating accurate descriptions of a fabric material given an image of it. Similarly to the retrieval application, we target fine-grained description, with precise properties described, which are not limited to high-level semantics. This further allows us to explicitly observe the ingestion of the concepts stemming from our dataset by Large Language Models.

\subsubsection{Implementation}
\label{subsubsec:implementation-blip}

We leverage and fine-tune BLIP~\cite{BLIP} for caption generation, and process our data as described in Section~\ref{subsubsec:implementation-clip}; however, in this case we do not split the sentences, ensuring full descriptions are seen by the model. We fine-tune the generative head of BLIP, starting from 
what we term \emph{native} BLIP: the VIT-B/16 model pre-trained on 129M images from LAION + CapFilt-L (model\_base\_capfilt\_large). We train the network for 12 epochs using the Adam optimizer with weight decay regularization using a decay parameter of $0.05$, an initial learning rate of $1e^{-5}$ and a batch size of $24$. The minimum number of generated tokens is set to $5$, and the maximum to $80$. This takes approximately 9 hours to train on a single RTX3090.
Once fine-tuned, we use nucleus sampling for tokens\new{~\cite{holtzman2020nucleus}}, letting us generate varied descriptions for each image.

 \subsubsection{Keyword Extraction}
 
 While we mainly focus on generating natural language descriptions (using the model described in Section~\ref{subsubsec:implementation-blip}), simple keywords can also be convenient in several search or classification scenarios. 
 Therefore, using our understanding of fabric descriptions in terms of common lexicon, main attributes and structure (Section~\ref{sec:dataset-analysis}), we automatically extract keywords for an image based on the generated descriptions. For that, we use the first five descriptions of an image, post-process the text as explained in Section~\ref{subsubsec:postpro}, extract a set of keywords per attribute from our lexicon, and order them by importance (number of descriptions in which they appear) and the rank product of their attribute. Resulting keywords for real images are shown in Figure~\ref{fig:application:descriptiongen_real}, and \fillin{we include automatically extracted keywords for all material samples as part of our text2fabric dataset}.

\subsubsection{Results}

We show here description results from our fine-tuned BLIP model, 
together with comparisons to native BLIP.

\begin{figure*}
 \centering
 \includegraphics[width=\textwidth]{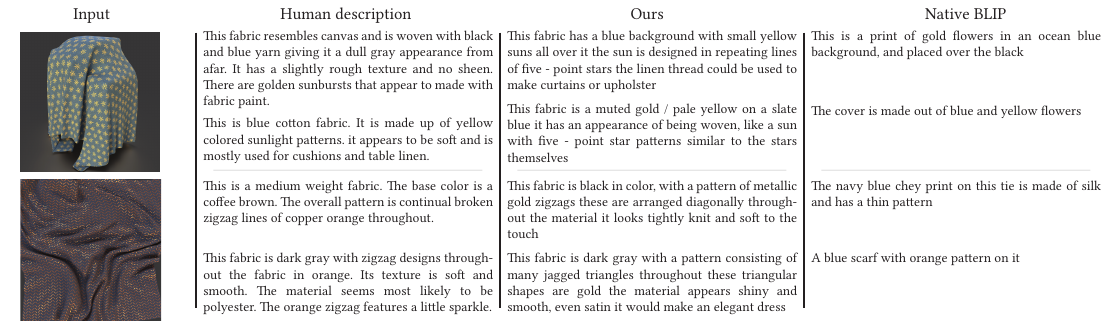}
 \caption{
 Description generation results for synthetic materials from the test set. We show results on two different geometries: \emph{baseline} (\emph{top}), seen during fine-tuning, and \emph{plane\_draped} (\emph{bottom}), unseen during fine-tuning. The descriptions included are, \emph{from left to right}: from our gathered dataset, provided by humans; generated by our fine-tuned model; and generated by native BLIP.
 Our descriptions not only are closer in style to human descriptions, but are also better at conveying fine-grained appearance and details.
 }
  \label{fig:application:descriptiongen_synth}
 \end{figure*}

Figure~\ref{fig:application:descriptiongen_synth} shows captioning results on synthetic images, with two different geometries, and materials from the test set (unseen during training). 
For each fabric sample, we include: descriptions from our dataset, provided by humans; descriptions generated by our fine-tuned model; and descriptions generated by native BLIP.
We observe how our fine-tuned model generates descriptions that better convey fine-grained material appearance, are more accurate and with more attention to detail, and match more closely the style of human descriptions.
 
\begin{figure*}
 \centering
 \includegraphics[width=\textwidth]{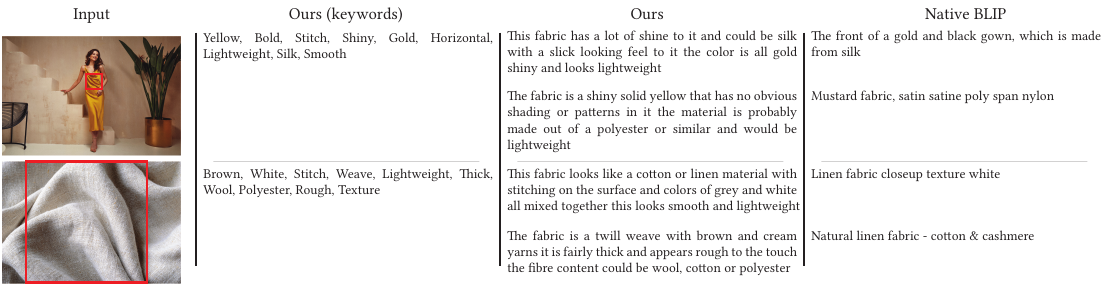}
 \caption{
 Description generation results on real images (the input is marked by a red square) for both our fine-tuned model and native BLIP.
 We also include results of our automatic keyword extraction. 
 We can see that our fine-tuned model for caption generation generalizes well to real photographs.}
 \label{fig:application:descriptiongen_real}
 \end{figure*}

We further demonstrate our results on real images 
containing fabrics in Figure~\ref{fig:application:descriptiongen_real}. We crop the fabric area of interest (marked by a red square), and generate descriptions for it. The descriptions generated using our fine-tuned model contain significantly richer information than the native BLIP results, trained only on general internet images and high-level descriptions. Additionally, our keyword extraction method is capable of automatically extracting relevant keywords from our generated sentences, which can be useful for, e.g., automatic tagging. These results further show that the fine-tuning on our high quality renderings generalizes well to real photographs.

\subsection{Invariance of the Latent Space to Geometry and Illumination}
\label{subsec:apps:invariance}
Our dataset significantly helps to improve the invariance to lighting and geometry of large vision-language models representations. We illustrate this by fine-tuning CLIP (as described in Section~\ref{subsubsec:implementation-clip}) using all the renderings of our materials, with five different geometries and three environment illuminations (\emph{baseline}, \emph{outdoor}, \emph{studio}) associated to our \nbvaliddescription descriptions. While we could use contrastive learning to try to learn an invariant representation instead, the only available supervision would be whether or not two images show the same material, making the creation of a perceptually smooth representation challenging.
Using our descriptions as anchor contributes to a smooth representation space, 
allowing the model to learn a more robust notion of material appearance than that of native CLIP, as shown by our evaluation, described next.

\begin{table}
 \caption{
 Average cosine similarity (and associated standard deviation) between pairs of images exhibiting: 
 the same material (and lighting conditions) and different geometries (\emph{varying geometry}); and 
 the same material (and geometry) and different lighting conditions (\emph{varying lighting}).
 Our fine-tuned representation (\emph{first row}) finds images of the same material to be more similar, despite geometry or lighting variation, than native CLIP space (\emph{second row}). \new{We include results of the Wilcoxon signed-rank test showing effect size | p-value (\emph{bottom row}). Differences are statistically significant and with large effect sizes for both experiments.}}
\begin{tabular}{r|c|c|}
 & Varying Geometry & Varying Lighting \\ \hline
 Ours & $0.951 \pm 0.012$ & $0.973 \pm 0.007$ \\
 Native CLIP & $0.835 \pm 0.042$ & $0.945 \pm 0.011$ \\ \hline
 \new{Wilcoxon signed-rank} & 0.866 | <0.0001 & 0.846 | <0.0001 \\ \hline
\end{tabular}
 \label{tab:application:cross_geom_light}
\end{table}

\begin{figure}
    \centering
    \includegraphics[width=\columnwidth]{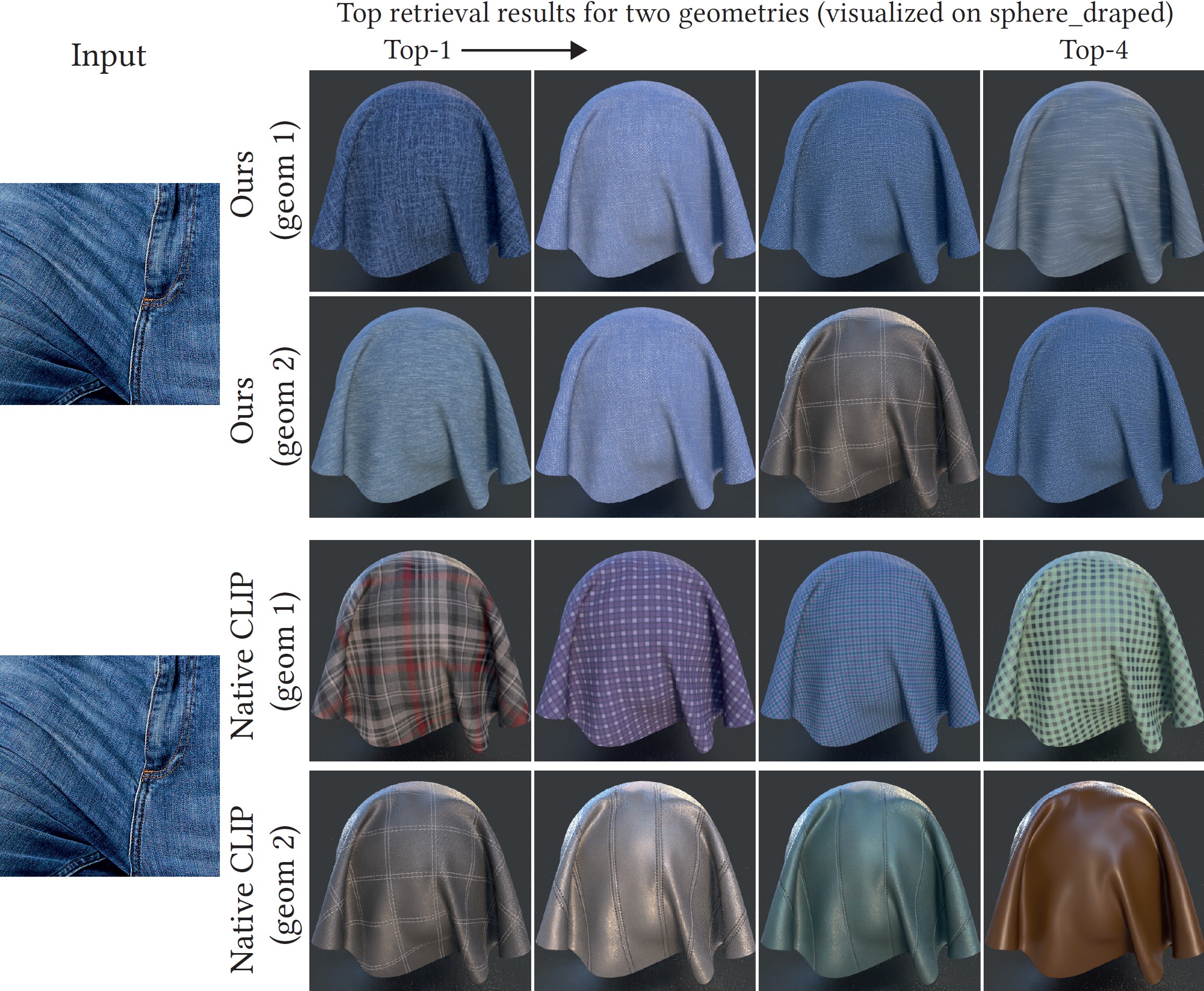}
    \caption{
    Latent space invariance to geometry. 
    Top-4 results of image-based search in databases rendered on different geometries (geom 1: \emph{sphere\_draped}; geom 2: \emph{plane}; see text for details). 
    We display all results rendered on \emph{sphere\_draped} for easier comparison.
    Our representation is significantly less affected by the geometry than the latent space of native CLIP, learning a more precise notion of material appearance. }
    \label{fig:application:realImageRetrievalDifferentGeometry}
\end{figure}

\begin{figure}
    \centering
    \includegraphics[width=\columnwidth]{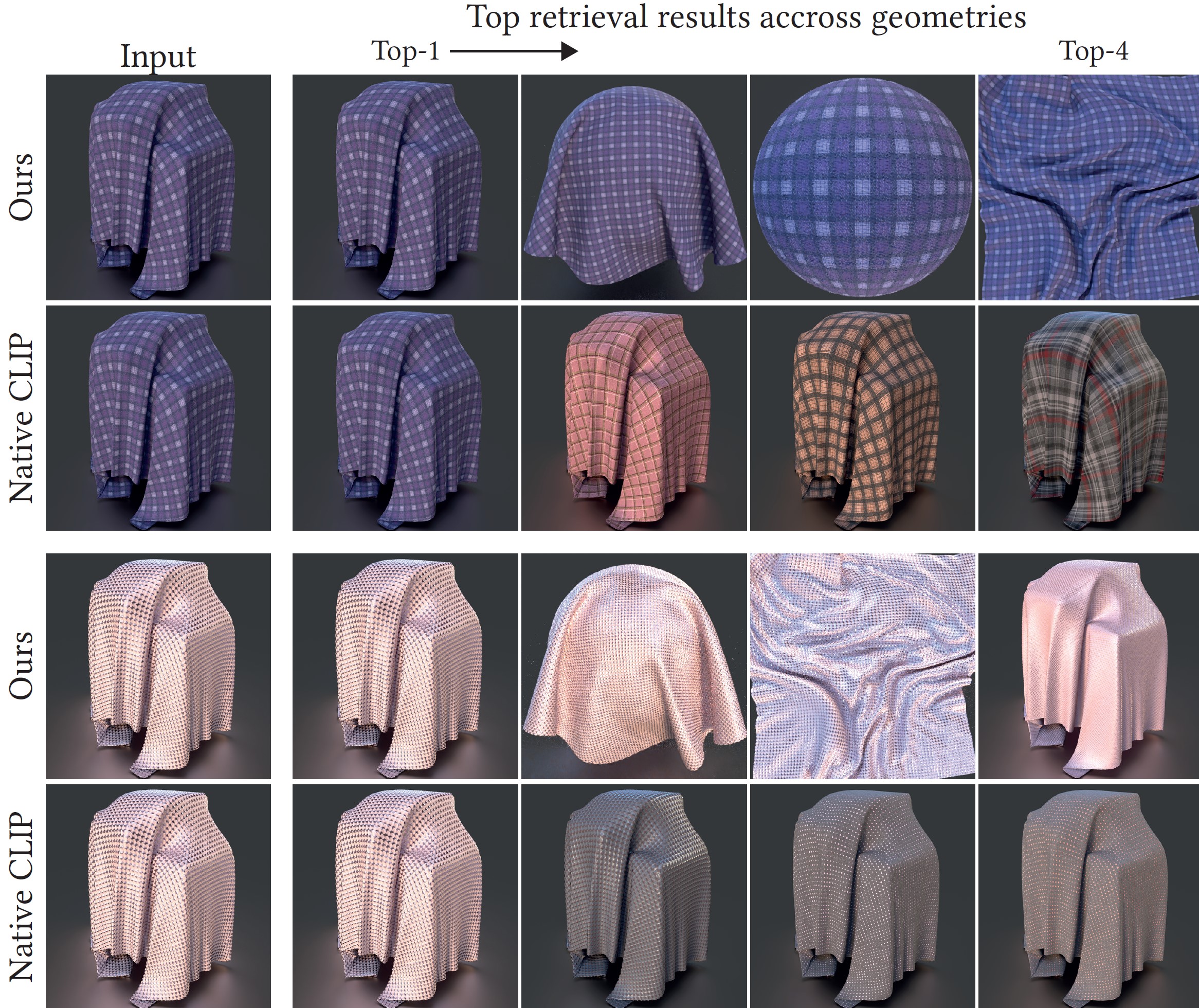}
    \caption{
    Top-4 results of image-based search in a database containing all our geometries: We see that native CLIP is heavily biased by the geometry of the input image, while our fine-tuned model focuses on the material appearance and is capable of recovering the same material across geometries.
    }
    \label{fig:application:crossGeometryRetrieval}
\end{figure}

In Table~\ref{tab:application:cross_geom_light} (first column, \emph{varying geometry}) we evaluate the average cosine similarity between pairs of images rendered with the same material and lighting but different geometries, computed both in native CLIP space and in the latent space of our fine-tuned model. Our representation is, on average, more invariant to geometry than the original CLIP features.
The same evaluation for pairs of images rendered with the same material and geometry but different lighting conditions (Table~\ref{tab:application:cross_geom_light}, \emph{varying lighting}), shows a similar trend, although less pronounced. In both cases, the lower standard deviation between pairwise similarities using our representation suggests a greater stability across variations. \new{A Wilcoxon signed-rank test shows that these differences are statistically significant (p-value<0.0001), with effect sizes considered large for both geometry and lighting variations~\cite{rosenthal1994parametric}.}

We qualitatively evaluate this property in Figure~\ref{fig:application:realImageRetrievalDifferentGeometry}: we assess whether, given a real photograph as input to an image-based search (see Section~\ref{subsec:apps:search}), the results change depending on the geometry present in the database we search in (for this test, each database we search in has all materials rendered with a single geometry). The figure shows results for the search in the \emph{sphere\_draped} and \emph{plane} geometries databases, and we display all results rendered on \emph{sphere\_draped} for easier comparison. We can see that our representation is significantly more consistent than native CLIP on varying geometries, and better at learning features at material scale.

We further pursue this evaluation in Figure~\ref{fig:application:crossGeometryRetrieval}.
Here, we seek at retrieving a given \emph{test} (i.e., unseen during training) material rendered on a given geometry, performing image-based search in a database containing renderings of all 3,000 materials and five geometries. As expected, with both native CLIP and our fine-tuned model, the first result is the same material and geometry. However, it is clearly apparent that the native CLIP representation is heavily biased by the geometry in the input image, while our representation better identifies the same material across geometries.

\vspace{-0.45em}
\section{Discussion and Future Work}
\label{sec:discussion}

We have presented text2fabric, a comprehensive, large-scale public 
dataset relating the visual appearance of fabrics to natural language. We have analyzed and curated a rich lexicon, classifying it into eleven attributes and highlighting key concepts used by humans when describing fabrics. We have further proposed several applications including fine-grained retrieval, image-based search, and caption generation, and shown how foundational, state-of-the-art vision-language models such as CLIP~\cite{CLIP} or BLIP~\cite{BLIP} struggle to represent fine-grained concepts of appearance, unless fine-tuned on our dataset.

\begin{figure}
    \centering
    \includegraphics[width=\columnwidth]{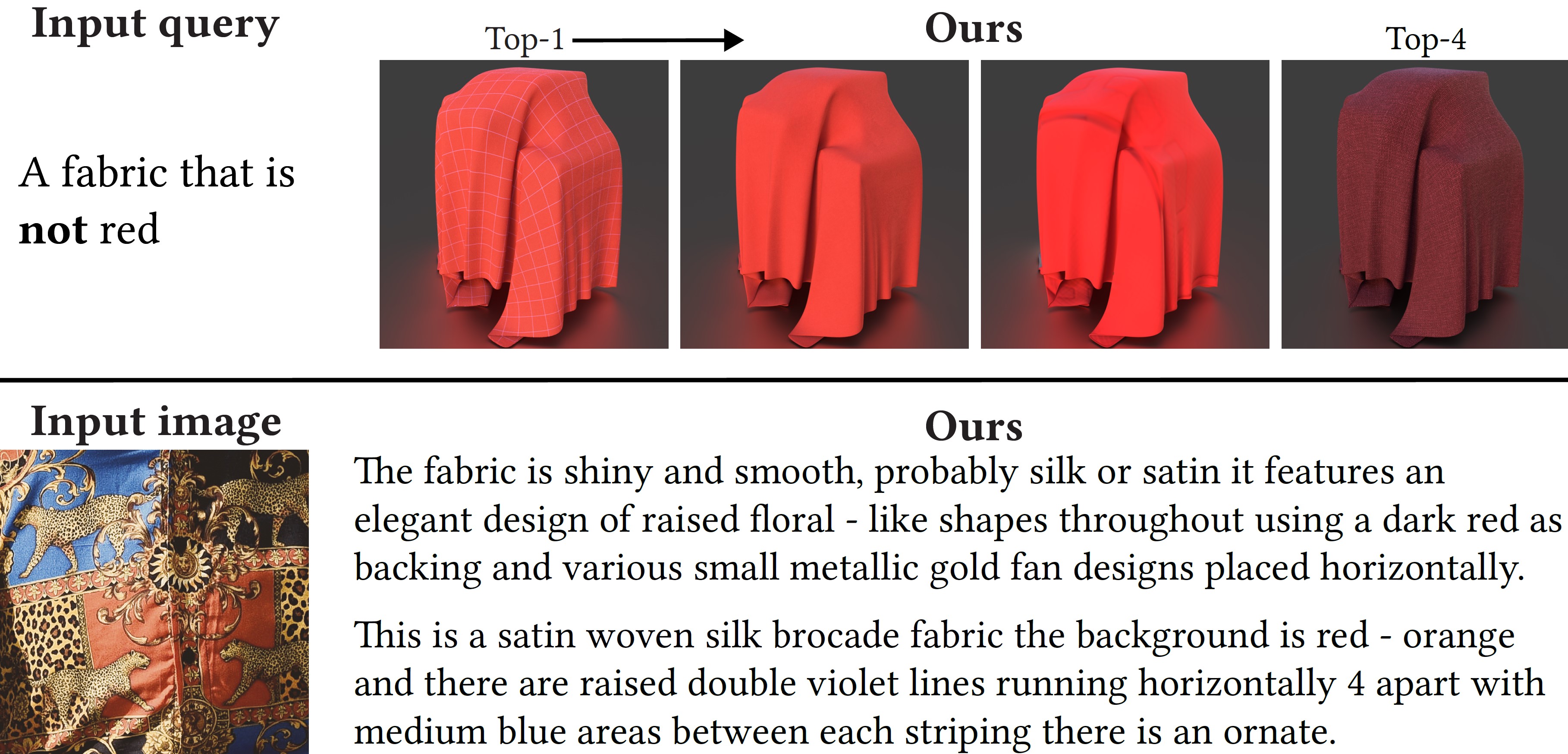}
    \caption{
    Limitations. \emph{Top row:} The text-based fine-grained retrieval does not always work well for negative queries that people do not use when describing (e.g., one would not say ``this is a non-red fabric'' or ``this fabric is not red''). \emph{Bottom row:} While our generated descriptions capture many relevant details, the intricacy of the pattern image results in the model missing some features that are salient to humans, such as the leopards.
    }
    \label{fig:limitations}
\end{figure}

Our work is not free of limitations. First, as in all studies, our results are only strictly valid for our particular set of stimuli; for example, in the fabric samples used to generate our data, military characteristics and some weathering features  are highly correlated, such as camouflage and dirt, as shown in Figure~\ref{fig:dataset:attributes} (right). This leaves the door open for future extensions of our dataset to explore these correlations further. 
\new{Second, we decided to choose non-expert describers (albeit familiar with fashion or design), to target a wider audience for our applications, given that experts usually rely on highly specialized concepts, difficult to understand by the general public. This might lead to the descriptions including some inaccuracies (e.g., due to uncertainty in the fabric type), or common misunderstandings about cloth (e.g., confusing ``stitching'' with ``weaving''). While these are a reflection of assumptions and biases from common users, it could be a limitation in certain scenarios, e.g., involving experts.}
\new{Additionally,} as expected, 
some text-based queries are not fully understood by our models, as shown in Figure~\ref{fig:limitations} (top row). Fabric samples are typically not described in terms of \emph{not} having a certain characteristic (people do not say ``this is not a red fabric'' or ``this fabric is not red''); as a result, our fine-tuned models (and their native counterparts) struggle with such queries. Another limitation regarding caption generation occurs in the presence of very complex, intricate designs, where the descriptions produced may fail to capture certain aspects that would be prominent for a human. An example can be seen in Figure~\ref{fig:limitations} (bottom row), where despite the richness of the generated caption, it fails to mention the presence of leopards in the fabric. 

Our work opens up exciting avenues for future research, which we describe in the following paragraphs. 

\paragraph{Generalization} An interesting question from our work is the exploration of how well our methodology generalizes beyond fabrics, maintaining a similar intra-class variation description quality. We argue that our methodology can be readily applied to other material datasets and classes, including both data gathering and analysis, which could in turn enable similar applications to the ones described in Section~\ref{sec:applications}. As a proof-of-concept, and without incurring in the cost of gathering a whole new corpus of descriptions, we resort to Adobe Stock~\shortcite{AdobeStock}, a popular service where assets can be searched by class, and are tagged with keywords provided by artists. We gather the keywords corresponding to four material classes (``wood'', ``stone'', ``brick'' and ``metal''), quite different in nature from fabrics; while Adobe Stock does not provide free text descriptions, 
we aim to assess 
to what extent the keywords are well represented by our attributes found in Section~\ref{subsec:data-analysis-attributes}. 

\begin{table}
\caption{Precision (ratio between true positives and predicted positives) of the automatic classification of keywords from other material categories into our generic attributes. 
}
\scriptsize
\begin{tabular}{l|c|c|c|c|c|c|c|c|}
\cline{2-9}
                            & color & lightness & metallic & pattern & touch & use   & weath. & mat\_type \\ \hline
\multicolumn{1}{|l|}{wood}  & 0.71  & 1.00      & 1.00     & 0.68    & 0.75  & 0.84  & 1.00   & 0.89      \\
\multicolumn{1}{|l|}{brick} & 0.72  & 0.80      & 0.67     & 0.59    & 0.68  & 0.83  & 1.00   & 0.75      \\
\multicolumn{1}{|l|}{stone} & 0.52  & 0.95      & 0.65     & 0.52    & 0.75  & 0.85  & 0.83   & 0.95      \\
\multicolumn{1}{|l|}{metal} & 0.53  & 0.47      & 0.70     & 0.68    & 0.75  & 0.88  & 1.00   & 0.67      \\ \hline
\multicolumn{1}{|l|}{avg}   & 0.62  & 0.80      & 0.76     & 0.62    & 0.73  & 0.85  & 0.96   & 0.82      \\ \hline
\end{tabular}
\label{tab:dataset:generalization}
\end{table}

We remove the attributes that are specific to fabrics, namely: \emph{sewing}, \emph{weight} and \emph{military}, and rename \emph{fabric\_type} with the corresponding material category (e.g., \emph{wood\_type}).
We then automatically classify keywords from all four classes into the attributes (see Section~\ref{subsec:data-analysis-attributes}). Table~\ref{tab:dataset:generalization} shows precision values for each attribute and class, i.e., how many keywords assigned to the attribute truly belong to it (we obtain the ground truth by manual classification). We see how precision values are reasonably high, suggesting generality of our attributes. The exceptions are \emph{color}, whose low precision is due to the presence in our lexicon of common objects used as colors (e.g., olive), and \emph{pattern}, probably due to the very general nature of this attribute. While this is a very preliminary analysis, we believe it hints at the generalization capabilities of our methodology and derived attributes, and may inspire future work in this direction. 

Another interesting avenue of research is exploring generalization \textit{beyond} material categories, such as video or meshes. Besides, since our methodology lets us relate synthetic graphics primitives to natural language, we are then free to use our primitives under arbitrary conditions, for example adapting them to a specialized context such as garments, or specific environments. An interesting line of future work would be to exponentially augment datasets by combining geometries and materials descriptions into new complete descriptions of the combination, enabling virtually infinite geometry, environment and material combinations for downstream natural language and visual tasks.

\vspace{-1.2em}
\new{\paragraph{Dataset extension} 
We used rendered images instead of photographs due to the large size of our dataset, since capturing 45,000 samples of different fabrics under controlled, professional conditions would impose non-negligible costs. On the other hand, using existing photographs would introduce uncontrolled variations in geometry and lighting, which may hamper the task of describing material appearance. Nevertheless, carefully augmenting our dataset with real images could enhance the performance of some applications. 
Additionally, our dataset could be further extended by adding expert terminology to the textual data, and used for instance to investigate social associations typically derived from clothing, such as occupation, personality, or socioeconomic status.} 

\paragraph{Generative models} While the challenging task of material generation is out of the scope of this study, recent material generation models have used different images as conditions~\cite{Zhou22, Guo20}. As shown, our dataset enables better visual correspondence between appearance and natural language. Combined with the strong prior of a fabric material generation model, our dataset could significantly improve text-conditioned material generation and editing.

\paragraph{Physical properties} 
Our dataset consists only of static stimuli. Although it has been shown that visual appearance dominates over dynamics when describing most fabrics, certain characteristics may be better conveyed by simulating the physics of such fabrics in motion~\cite{aliaga2015sackcloth}. Exploring the relative weights of appearance and dynamics on the perception of fabrics is an interesting research topic, although requiring a significant amount of work to model and simulate the physics of the fabrics.  

We hope that text2fabric helps enable these and other studies, which in turn may lead to the creation of novel applications. 
\section*{Acknowledgments}
We thank Kushal Kafle for his help in navigating the NLP domain; Yulia Gryaditskaya and Zoya Bylinskii for advice about data collection and user studies; Jorge Gracia del Rio for NLP pointers; and Sergio Izquierdo and the members of the Graphics and Imaging Lab 
for discussions and help preparing the final figures.
We would also like to thank the Substance 3D Assets team for the creation of the materials, all the participants in the user studies, and the anonymous reviewers.
This work has received funding from the European Research Council (ERC) within the EU's Horizon 2020 research and innovation programme (project CHAMELEON, Grant No. 682080) and under the Marie Skłodowska-Curie grant agreement No. 956585 (project PRIME), as well as a generous donation from Adobe. Julia Guerrero-Viu was supported by the FPU20/02340 predoctoral grant.

\bibliographystyle{ACM-Reference-Format}
\bibliography{bibliography}
\end{document}